\documentclass[journal]{IEEEtran}

\usepackage{multicol}
\usepackage{amsfonts}
\usepackage{amssymb}
\usepackage{stfloats}
\usepackage{cite}
\usepackage{graphicx}
\usepackage{epstopdf}
\usepackage{psfrag}
\usepackage{diagbox}
\usepackage{subfigure}
\usepackage{amsmath}
\usepackage{array}
\usepackage{stfloats}
\usepackage{multirow}
\usepackage{color}
\usepackage{booktabs}
\usepackage{url}
\usepackage{graphicx}
\usepackage{enumitem}
\usepackage{lipsum} % 调用跨栏长公式宏包
\usepackage{float}
\usepackage{marvosym} 
\usepackage{url}
\usepackage{makecell}
\usepackage{bm}
\usepackage{eso-pic}
\usepackage{booktabs} % For better looking tables
\usepackage{multirow} % To combine cells vertically
\usepackage{subcaption}
\usepackage[linesnumbered,ruled,vlined]{algorithm2e}
\usepackage{cases}

\newtheorem{Thm}{Theorem}
\newtheorem{Lem}{Lemma}

\newtheorem{Prob}{Problem}

\begin{document}

\title{A Model-Data Dual-Driven Resource Allocation Scheme for IREE Oriented 6G Networks}

\author{Tao~Yu, Simin~Wang, Shunqing~Zhang,~\IEEEmembership{Senior Member, IEEE}, Xiaojing~Chen,~\IEEEmembership{Member, IEEE}, Zi~Xu, Xin Wang,~\IEEEmembership{Fellow, IEEE}, Jiandong Li,~\IEEEmembership{Fellow, IEEE}, Junyu Liu,~\IEEEmembership{Member, IEEE} and Sihai Zhang,~\IEEEmembership{Senior Member, IEEE}
\thanks{This work was supported by the National Key Research and Development Program of China under Grants 2022YFB2902304, and the Science and Technology Commission Foundation of Shanghai under Grants 24DP1500500, and 24DP1500700.}
\thanks{Tao Yu, Simin Wang, Shunqing Zhang and Xiaojing Chen are with Shanghai Institute for Advanced Communication and Data Science, Key laboratory of Specialty Fiber Optics and Optical Access Networks, Shanghai University, Shanghai, 200444, China (e-mails: \{yu\_tao, siminwang, shunqing, jodiechen\}@shu.edu.cn).

Zi Xu is with the Department of Mathematics, Shanghai University, Shanghai, 200444, China (e-mail: xuzi@i.shu.edu.cn).

Xin Wang is with the Key Laboratory for Information Science of Electromagnetic Waves (MoE), Department of Communication Science and Engineering, Fudan University, Shanghai 200433, China (e-mail: xwang11@fudan.edu.cn).

Jiandong Li and Junyu Liu are with the State Key Laboratory of Integrated Service Networks, Institute of Information Science, Xidian University, Xi’an 710071, China (e-mail: jdli@mail.xidian.edu.cn; junyuliu@xidian.edu.cn).

Sihai Zhang is with the Key Laboratory of Wireless-Optical Communications, Chinese Academy of Sciences University of Science and Technology of China, Hefei 230026, China (e-mail: shzhang@ustc.edu.cn). }
\thanks{Corresponding Author: {\em Shunqing Zhang}.}
}

\markboth{Journal of \LaTeX\ Class Files,~Vol.~14, No.~8, August~2015}%·······
{Shell \MakeLowercase{\textit{et al.}}: Bare Demo of IEEEtran.cls for IEEE Journals}

\maketitle

\begin{abstract}
The rapid and substantial fluctuations in wireless network capacity and traffic demand, driven by the emergence of 6G technologies, have exacerbated the issue of traffic-capacity mismatch, raising concerns about wireless network energy consumption. To address this challenge, we propose a model-data dual-driven resource allocation (MDDRA) algorithm aimed at maximizing the integrated relative energy efficiency (IREE) metric under dynamic traffic conditions. Unlike conventional model-driven or data-driven schemes, the proposed MDDRA framework employs a model-driven Lyapunov queue to accumulate long-term historical mismatch information and a data-driven Graph Radial bAsis Fourier (GRAF) network to predict the traffic variations under incomplete data, and hence eliminates the reliance on high-precision models and complete spatial-temporal traffic data. We establish the universal approximation property of the proposed GRAF network and provide convergence and complexity analysis for the MDDRA algorithm. Numerical experiments validate the performance gains achieved through the data-driven and model-driven components. By analyzing IREE and EE curves under diverse traffic conditions, we recommend that network operators shall spend more efforts to balance the traffic demand and the network capacity distribution to ensure the network performance, particularly in scenarios with large speed limits and higher driving visibility.
\end{abstract}


\section{Introduction} \label{sect:intro}

The emergence of advanced applications, including extended reality (XR), the tactile Internet and autonomous vehicles, presents new challenges to the sustainable development of sixth-generation (6G) wireless networks \cite{zhang20206g,jiang2021road}. Different from previous generations of wireless networks, 6G is promising to have highly dynamic traffic variations in both temporal and spatial domains, and leads to capability mismatches and performance degradations, as explained in \cite{bor2016new,yang2017proactive,9627726}. Since the sustainability issue has been listed as a key performance requirement for 6G, several innovative approaches have been presented. For instance, integrating millimeter-wave and terahertz communication in ultra-dense networks can yield more than $50\%$ energy efficiency (EE) gains \cite{10018462}. Intelligent reflecting surface (IRS)-based resource allocation methods have been shown to achieve $300\%$ improvement in EE \cite{huang2019reconfigurable}. 

All the aforementioned energy efficient schemes, however, rely on the assumption that the corresponding traffic variations are stable and the traffic dynamics are rarely considered. To address this problem, the Lyapunov queue is used in \cite{kim2017traffic} to capture spatial-temporal mismatch between traffic and offered data rate. By combining the Hungarian algorithm with the traffic prediction technology, \cite{yang2017proactive} provides a proactive drone deployment strategy for the traffic-capacity mismatch caused by flash crowd traffic in 5G networks and maximized the service capability of the drone cells. Recently, some data-driven schemes such as Deep Reinforcement Learning (DRL) and Long Short-Term Memory (LSTM) have emerged as powerful tools to deal with the aforementioned mismatch and maximize the EE performance of the wireless systems \cite{chen2021deep, 9611038}.

%However, existing resource allocation schemes mainly have following two limitations. First, these schemes are designed for the conventional EE metrics which are unable to integrate the  traffic-capacity mismatch, thus potentially leading to suboptimal results \cite{10605762}. Second, existing schemes either rely heavily on the high-precision mathematical models or require comprehensive traffic profiles for effective learning and prediction, limiting their practical applicability. In our previous work \cite{yu2022novel}, a novel EE metric named Integrated Relative Energy Efficiency (IREE) is proposed to incorporate the traffic-capacity mismatch in the EE evaluation via the famous Jensen-Shannon (JS) divergence \cite{manning1999foundations}, and provides a more comprehensive assessment of energy efficiency.

% 这里写一下大家通过各类不同的EE指标来解决mismatch的问题，提出解决问题的新思路，以及存在的主要问题。

In this paper, we target to provide design guidelines for energy efficient 6G networks under \emph{spatial-temporal mismatch}, and propose a model-data dual-driven resource allocation (MDDRA) framework to deal with the traffic-capacity mismatch problem. Specifically, we apply a model-driven Lyapunov queue based long-term correction to assemble the historical information and a data-driven short-term traffic prediction network named Graph Radial bAsis Fourier (GRAF) to characterize the real-time traffic variations under incomplete traffic knowledge. Through this dual-driven approach, the proposed scheme is able to adapt to the spatial-temporal traffic variations and recursively optimize the wireless resources for performance maximization. 

\subsection{Related Works} \label{subsect:related}

The traffic-capacity mismatch problem has been widely discussed in the existing literature, which includes the spatial mismatch \cite{10605762}, the temporal mismatch \cite{10460374}, and the spatial-temporal mismatch \cite{9888068}. When the mismatch happens, some conventional resource allocation schemes need to track this mismatch and optimize the resources to fill in the gap, which belongs to the time coupled optimization (TCO) problem as specified in \cite{sun2025online}. 

By applying the TCO formulation, many {\em model-driven} resource scheduling schemes have been discussed. For instance, the network deployment problem has been formulated as a mixed integer non-linear problem to solve the spatial mismatch in \cite{chiaraviglio20215g}, and the Lyapunov framework has been adopted to deal with the temporal mismatch in \cite{10460374}. If spatial-temporal mismatch is involved, a Stackelberg game-based approach has been discussed in \cite{9888068} to coordinate the deployment of BS enabled by unmanned aerial vehicles and carry out radio resource allocation accordingly. In addition to model-driven approaches, {\em data-driven} resource allocation schemes have also been investigated during recent years. As an example, we can leverage contextual datasets such as geospatial, spatial-temporal traffic, and the demographic data to reduce the mismatch in the spatial domain \cite{haile2020data}. In the temporal domain, DRL based schemes can leverage extensive historical channel or traffic data to dynamically adjust resource allocation policies \cite{9372298}, while historical drone position can also be utilized by this methodology to address spatial-temporal mismatch\cite{khurshid2023drl}.

The above solutions are based on the assumption that all historical traffic and capacity information is available before the algorithm execution, which is generally difficult to satisfy in many real-world applications. With incomplete traffic and capacity information, recent efforts deal with this issue by extending traditional solutions for utility-maximization or energy-minimization TCO problems. For example, model-driven approaches using Lyapunov optimization or Hungarian algorithm-based online matching have been discussed in \cite{10.1145/3708893} and \cite{tong2016online}, and data-driven techniques, including DRL and LSTM-based predictors, have been proposed in \cite{7874185,9816080}. Since the adopted models and data sets are based on incomplete historical information, a model-data dual-driven framework has been proposed in \cite{10107793} to minimize the energy consumption as well.

\subsection{Contributions \& Organizations} \label{subsect:contribution}

In this paper, we deal with the traffic-capacity mismatch problem with incomplete historical information using Integrated Relative Energy Efficiency (IREE) metric \cite{yu2022novel}. Specifically, we propose an MDDRA scheme to deal with this TCO problem with incomplete information and improve the IREE performance by optimizing bandwidth and power allocation, and the main contributions are summarized as follows.

\begin{itemize}
\item{\em Dual-Driven Framework for Multi-Scale Mismatch.} 
In the proposed IREE optimized MDDRA framework, we apply a model-driven Lyapunov queue based long-term correction to assemble the historical mismatched information and a data-driven traffic prediction network to characterize short-term traffic variations. By applying this dual-driven framework, high-accurate models and complete spatial-temporal traffic information are no longer needed. This is because short-term traffic prediction errors can be compensated by Lyapunov queue based long-term correction and the rate stability conditions can be satisfied by specifically designed loss functions used in GRAF network.

% the GRAF network is trained to approximate it using the RBF layer. Since the FGO block focus on the dynamic of the network configurations, a complete traffic profile is not required for back propagation. During forward propagation, the GRAF architecture leverages the RBF layer output to effectively interpolate missing traffic data. This approach is justified by the universal approximation property of RBF layer and the spatial continuity of its output. Further, we have the following theorem.

\item{\em Data-Driven GRAF Network Design for Incomplete Traffic Profile}. Thanks to the universal approximation property of RBF layer and the spatial continuity of its output, the GRAF network proposed in this paper is able to focus on the complete historical network configuration profile rather than the incomplete traffic information. Guided by the model-driven long-term Lyapunov framework, a pre-training \& fine-tuning strategy is proposed to optimize the GRAF network for short-term traffic prediction. Starting from an initial point of minimal traffic-capacity mismatch, the proposed GRAF network gradually optimize the IREE metric by allocating the radio resources in wireless networks under incomplete traffic profile and the accumulated traffic-capacity mismatch errors due imperfect traffic prediction 
can be compensated by queueing constraints in the aforementioned Lyapunov framework.  

\item{\em Performance Comparison and Design Principle.} We compare the achievable IREE performance with several baselines to show advantages of the proposed MDDRA scheme. Based on some numerical examples, the proposed scheme can achieve 10.2\% to 20.7\% IREE gains, if compared with purely data-driven or model-driven approaches. In addition, we find several important design principles by extensive simulations and conclude that the network operator should allocate more power budget to ensure network services on roads with large speed limits and higher driving visibility.

\end{itemize}

The remaining part of this paper is organized as follows. We provide the system model in Section~\ref{sect:system_model} and formulate the IREE maximization problem in Section~\ref{sect:problem}, where we also provide the problem transformation. The IREE maximized MDDRA scheme is proposed in Section~\ref{sect:proposed_scheme}, with a detailed introduction of the proposed GRAF network. In Section~\ref{sect:num_res}, we provide some numerical examples on the proposed MDDRA scheme. Finally, concluding remarks are provided in Section~\ref{sect:conc}.

\section{System Model} \label{sect:system_model}

In this section, we briefly introduce the traffic and the network model, and give the definition of IREE for the dynamic wireless scenario.

\subsection{Traffic Model}

A unidirectional vehicular/pedestrian flow within geographic area $\mathcal{A}$ and time period $\mathcal{T}$ can be modeled by a convection-diffusion equation given as \cite{bonzani2000hydrodynamic},
\begin{eqnarray}
\label{eqn:convection-diffusion}
\partial_{\tau} \rho + \nabla_{\mathcal{L}} (\rho f (\rho)) = \nabla_{\mathcal{L}} \left ( g(\rho) \nabla_{\mathcal{L}} \rho \right ), \forall \mathcal{L} \in \mathcal{A}, \tau \leq \mathcal{T},
\end{eqnarray}
where $\rho(\mathcal{L}, \tau)$ is the normalized user density, $f (\rho)$ is the velocity field function and $g (\rho)$ is the diffusion function given by,
\begin{eqnarray}
\label{eqn:}
f (\rho) &=& (1- \rho^{\mu_1})^{\mu_2} V_{e}, \\
g (\rho) &=& \rho^{\vartheta_1} (1- \rho)^{\vartheta_2} T_{a}.
\end{eqnarray}
In the above equations, $V_{e}$ is the normalized equilibrium speed which reflects the flow speed when the system is in equilibrium. $T_{a}$ is the normalized anticipation time which reflects the time it takes to react to a situation after noticing it. $\mu_1, \mu_2, \vartheta_1, \vartheta_2 \geq 0$ are the empirical exponents. By selecting some appropriate parameters, equation \eqref{eqn:convection-diffusion} can be simplified into various classic models, such as the well-known follow-the-leader model \cite{tordeux2018traffic}, meanwhile, it can be easily extended to 2D scenarios with multiple flows in different directions \cite{berres2011adaptive}. 

The time-varying data traffic at the location $\mathcal{L}$ and the time stamp $\tau$ can be then given by \cite{yu2022novel},
\begin{eqnarray}
\label{def:traffic}
D(\mathcal{L}, \tau) &=&  \rho(\mathcal{L}, \tau) d (\mathcal{L}, \tau),
\end{eqnarray}
where $d (\mathcal{L}, \tau)$ is the wireless demands for the user at the location $\mathcal{L}$ and the time stamp $\tau$, and is given by the sinusoid superposition model as follows \cite{7277444},
\begin{eqnarray}
\label{eqn:}
d (\mathcal{L}, \tau) &=& \sum_i d^M_i(\mathcal{L}) \sin[\varphi_i(\mathcal{L}) \tau + \varphi^0_i(\mathcal{L})].
\end{eqnarray}
In the above equations, $\{ d^M_i(\mathcal{L}) \}$ are the amplitudes, $\{ \varphi_i(\mathcal{L}) \}$ and $\{ \varphi^0_i(\mathcal{L}) \}$ are the frequency components and initial phases of traffic variation, respectively\footnote{The above model are verified with measured traffic data in \cite{6gc4-y070-22} as discussed in\cite{folland2009fourier}.}.

Unfortunately, we don't have a priori knowledge of these traffic parameters. Instead, the users are encouraged to report their locations. Combining the location information and the traffic demand at the BSs, we can have the available traffic profile of last time stamp $\tau-1$, i.e.,
\begin{eqnarray}
\label{eqn:incomplete_traffic}
\left \{ \left [ \mathcal{L}_m^{\tau-1}, D_m^{\tau-1}    \right ]  \right \}_{m=1}^{N_U^{\tau-1}} &\subseteq& \left \{ \left [ \mathcal{L}, D(\mathcal{L}, \tau-1)   \right ]  \right \}_{\mathcal{L} \in \mathcal{A}},
\end{eqnarray}
where $N_U^{\tau-1}$ is the number of volunteer users, $\mathcal{L}_m^{\tau-1}$ and $D_m^{\tau-1}$ are the location and the traffic demand of user $m$. As we mentioned before, this available traffic profile is characterized by inherent incompleteness across both temporal and spatial dimensions.

\begin{figure}[t] %[h]
\centering  
\includegraphics[height=6.5cm,width=8cm]{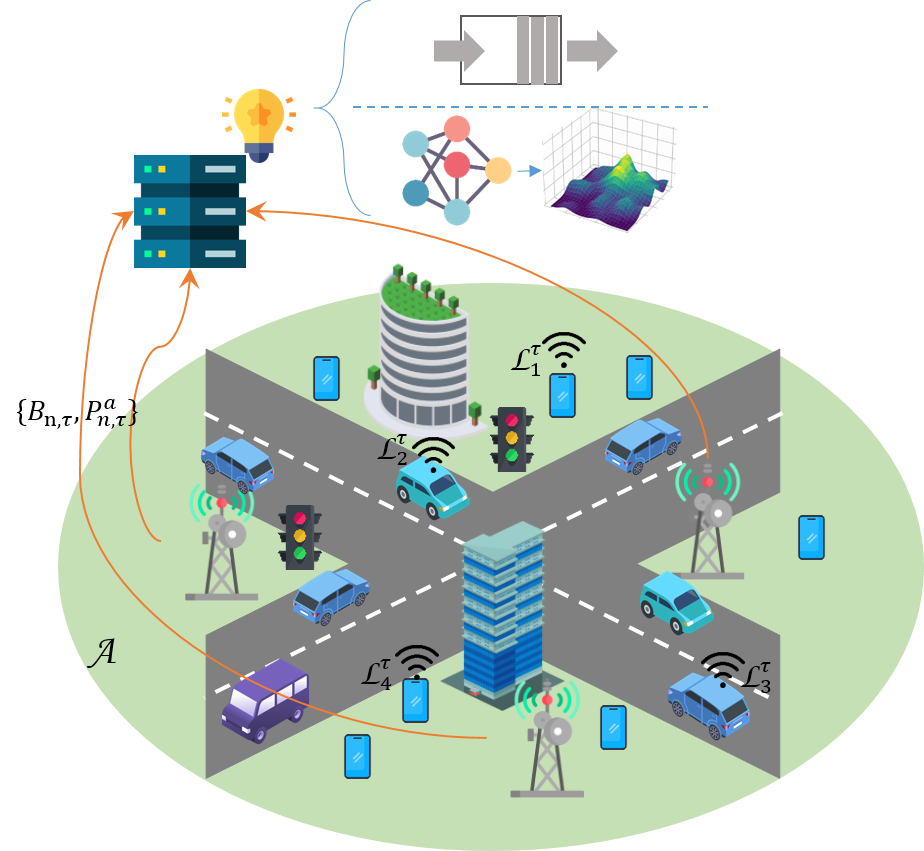}
\caption{An illustrative example of wireless networks and traffics within area $\mathcal{A}$ at time $\tau$, where there are only some volunteer users report their locations, and a dual-driven scheme is adopted to manage the resources of all BSs. }
\label{fig:scenario}
\end{figure}

\subsection{Network Model}

Consider a typical wireless communication network with $N_{BS}$ BSs as shown in Fig.~\ref{fig:scenario}, where the location, bandwidth, and transmit power of the $n^{th}$ BS at the time slot $\tau$ are denoted as $\mathcal{L}_n$, $B_{n,\tau}$, $P_{n,\tau }^{a}$, respectively. For any receiving entity at location $\mathcal{L}$, the equivalent capacity from the $n^{th}$ BS is provided by \cite{shannon1948mathematical},
\begin{eqnarray} 
\label{eqn:capacity}
C_n(\mathcal{L}, \tau) &=& B_{n,\tau} \log_2 \left (1 + \frac{ P_{n,\tau}^{a}/L(\mathcal{L}, \mathcal{L}_n)}{ \sigma^2 B_{n,\tau}} \right),
\end{eqnarray}
where $\sigma^2$ represents the power spectrum density of the additive white Gaussian noise (AWGN). The normalized path loss coefficients relative to the $n^{th}$ BS, $L(\mathcal{L}, \mathcal{L}_n)$, are derived from \cite{ku2013spectral},
\begin{eqnarray} 
\label{eqn:pathloss}
L(\mathcal{L}, \mathcal{L}_n) = \gamma \Vert\mathcal{L} - \mathcal{L}_n\Vert_2^{\alpha} + \beta, \quad \forall n \in [1,\ldots, N_{BS}].
\end{eqnarray}
In the above expression, $\alpha \geq 2$ denotes the path loss exponent and $\beta, \gamma > 0$ are the normalization factors.

Summing over all $N_{BS}$ BSs, the total capacity $C_{T}(\mathcal{L}, \tau)$ and the total power consumption $P_T (\tau)$ are defined by,
\begin{eqnarray}
C_{T}(\mathcal{L}, \tau) & = & \sum_{n=1}^{N_{BS}} C_n(\mathcal{L}, \tau), \label{eqn:tot_cap} \\
\label{eqn:total_pow_def}
P_T (\tau) & = & \sum_{n=1}^{N_{BS}} \left[\lambda P_{n,\tau}^{a} + P^{c}\right],
\end{eqnarray}
where $\lambda$ signifies the amplification coefficient linked to the power amplifier efficiency and $P^{c}$ indicates the static circuit power.

To quantify the statistical mismatch between network capacity and traffic distribution, we introduced a novel energy efficiency metric called integrated relative energy efficiency (IREE) in prior research \cite{yu2022novel}. This metric integrates non-uniform traffic patterns into EE evaluation through the established Jensen-Shannon (JS) divergence \cite{manning1999foundations}. Specifically, the definition of IREE is given by:
\begin{eqnarray} 
\label{eqn:def_iree}
\eta_{IREE}(C_{T}, D) =  \frac{\min\{C_{Tot}^{\mathcal{T}},D_{Tot}^{\mathcal{T}}\}\left[1 - \xi^{\mathcal{T}} \right]}{ P^{\mathcal{T}}_{Tot} },
\end{eqnarray} 
where $P_{Tot} = \int_{\mathcal{T}} P_T (\tau)\textrm{d}\tau$, $C_{Tot}^{\mathcal{T}} = \int_{\mathcal{T}} \iint_{\mathcal{A}}C_{T}(\mathcal{L}, \tau) \textrm{d}\mathcal{L} \textrm{d}\tau$, and $D_{Tot}^{\mathcal{T}} = \int_{\mathcal{T}} \iint_{\mathcal{A}}D(\mathcal{L}, \tau) \textrm{d}\mathcal{L}\textrm{d}\tau$ denote the total power consumption, the overall network capacity and traffic requirements, respectively. $\xi ^{\mathcal{T}} = \frac{1}{2} \int_{\mathcal{T}} \iint_{\mathcal{A}}  \frac{C_{T}(\mathcal{L}, \tau)}{C_{Tot}^{\mathcal{T}}} \log_2 \left[ \frac{2 D_{Tot}^{\mathcal{T}} C_{T}(\mathcal{L}, \tau) }{ D_{Tot}^{\mathcal{T}} C_{T}(\mathcal{L}, \tau) +  C_{Tot}^{\mathcal{T}} D(\mathcal{L}, \tau) } \right] + \frac{D(\mathcal{L}, \tau)}{D_{Tot}^{\mathcal{T}}} \log_2 \left[ \frac{2  C_{Tot}^{\mathcal{T}} D(\mathcal{L}, \tau) }{ C_{Tot}^{\mathcal{T}} D(\mathcal{L}, \tau) + D_{Tot}^{\mathcal{T}} C_{T}(\mathcal{L}, \tau) } \right] \textrm{d}\mathcal{L} \textrm{d}\tau$ denote the JS divergence between network capacity and traffic requirement. Based on this definition, the IREE oriented schemes are able to consider both network capacity improvement and traffic-capacity mismatch simultaneously.

The following assumptions are made through the rest of this paper. First, the traffic $D(\mathcal{L}, \tau)$ is assumed to be ergodic over time. Second, the maximum transmit power and maximum total bandwidth are limited by $P_{\max}$ and $B_{\max}$, respectively. Last but not least, $\beta, \gamma, \lambda$, and $P^{c}$ are assumed to be constant during the evaluation period\footnote{The non-constant channel fading effects, such as shadowing, will be discussed through numerical results in Section~\ref{sect:num_res}.}.

\section{Problem Formulation and Transformation} \label{sect:problem} 

In this section, we formulate the IREE maximization problem and transform it into a feasible problem under complete spatial-temporal traffic profile in what follows.

\subsection{Problem Formulation}
\label{subsec:formu}

In order to obtain the optimal bandwidth and power allocation strategy, $\{B_{n,\tau}, P_{n,\tau}^{a}\}^{*}$, we formulate the IREE maximization problem as below.
\begin{Prob}[Original IREE Maximization] 
\label{prob:original_iree}
For a given target evaluation geographic area $\mathcal{A}$ and time period $\mathcal{T}$, the IREE of the entire wireless network with $N_{BS}$ BSs can be maximized by solving the following optimization problem.
\begin{eqnarray}
    \underset{\{B_{n,\tau}, P_{n,\tau}^{a}\}}{\textrm{maximize}} && \eta_{IREE}(C_{T}, D), \nonumber \\
    \textrm{subject to} && \eqref{eqn:convection-diffusion} - \eqref{eqn:def_iree}, \nonumber \\
    && \frac{\min\{C_{Tot}^{\mathcal{T}},D_{Tot}^{\mathcal{T}}\}\left[1 - \xi^{\mathcal{T}} \right]}{ D_{Tot}^{\mathcal{T}} } \geq \zeta_{\min}, \label{constraint:qos} \\
    && \sum_{n=1}^{N_{BS}} B_{n,\tau} \leq B_{\max}, B_{n,\tau} \geq 0, \label{constraint:max_bandwidth} \\
    && 0 \leq P_{n,\tau}^{a} \leq P_{\max}. \label{constraint:max_power} 
\end{eqnarray}
In the above mathematical problem, $B_{\max}$ and $P_{\max}$ denote the corresponding total bandwidth and transmit power limit, respectively. The constraint \eqref{constraint:qos} ensures the minimum traffic requirement with $\zeta_{\min} \in [0,1]$.
\end{Prob}

The inherent complexity of Problem~\ref{prob:original_iree} arises from several key challenges. Firstly, it resists decomposition into independent solutions at each time step due to a violation of the optimality principle \cite{sniedovich1978dynamic}. This violation stems from the strong temporal coupling introduced by the JS divergence $\xi ^{\mathcal{T}}$ and the fractional programming structure. Secondly, model-driven approaches prove inadequate due to the difficulty in obtaining an explicit expression for IREE. This difficulty arises from the lack of prior knowledge regarding traffic patterns ${D}(\mathcal{L},\tau )$ over the time period $\mathcal{T}$. Finally, even if some data-driven schemes are employed, the incomplete traffic profile \eqref{eqn:incomplete_traffic} will lead to substantial performance degradation under the existing traffic prediction \cite{bashir2018handling} or RL-based resource allocation schemes \cite{wang2019robust}.

\subsection{Problem Transformation}
\label{subsec:formu}

To circumvent the challenges posed by the incomplete traffic profile, this section focuses on a simplified version of the problem. This simplification allows for decoupling across time steps, enabling the application of standard Alternating Direction Method of Multipliers (ADMM) based schemes for efficient solution under complete traffic profile.

With some mathematical manipulations as given in Appendix~\ref{appendix:iree_lower_bound}, we have an IREE lower bound given as,
\begin{eqnarray}
\label{def:long_iree}
\eta^{\mathcal{T}}_{IREE} \triangleq \frac{(1 - \kappa) \sum^{\mathcal{T}}_{\tau=0} \min\{C_{Tot}(\tau),D_{Tot}(\tau)\} }{\sum^{\mathcal{T}}_{\tau=0} P_T (\tau)},
\end{eqnarray}
where $C_{Tot}(\tau) = \iint_{\mathcal{A}}C_{T}(\mathcal{L}, \tau) \textrm{d}\mathcal{L}$ and $D_{Tot}(\tau) = \iint_{\mathcal{A}}D(\mathcal{L}, \tau) \textrm{d}\mathcal{L}$ denote the total network capacity and traffic requirements at the time slot $\tau$. $\kappa$ is the cumulative JS divergence given by,
\begin{flalign}
&\kappa \triangleq \sum^{\mathcal{T}}_{\tau=0}\xi(\tau) = \sum^{\mathcal{T}}_{\tau=0}\frac{1}{2} \iint_{\mathcal{A}}  \frac{C_{T}(\mathcal{L}, \tau)}{C_{Tot}(\tau)} \times  \log_2 \notag \\
&\quad \Big[ \frac{2 D_{Tot}(\tau) C_{T}(\mathcal{L}, \tau) }{ D_{Tot}(\tau) C_{T}(\mathcal{L}, \tau)  +  C_{Tot}(\tau) D(\mathcal{L}, \tau) } \Big]  + \frac{D(\mathcal{L}, \tau)}{D_{Tot}(\tau)} \times  \notag \\
&\quad \log_2 \Big[ \frac{2  C_{Tot}(\tau) D(\mathcal{L}, \tau) }{ C_{Tot}(\tau) D(\mathcal{L}, \tau) + D_{Tot}(\tau) C_{T}(\mathcal{L}, \tau) } \Big]\textrm{d}\mathcal{L}, \label{def:kappa}
\end{flalign}
where $\xi(\tau)$ denotes the transient JS divergence. We replace the original IREE expression and reformulate Problem~\ref{prob:original_iree} as following.

\begin{figure*}[t]
\centering\includegraphics[height=6.5cm,width=18cm]%[width=0.9\textwidth]
{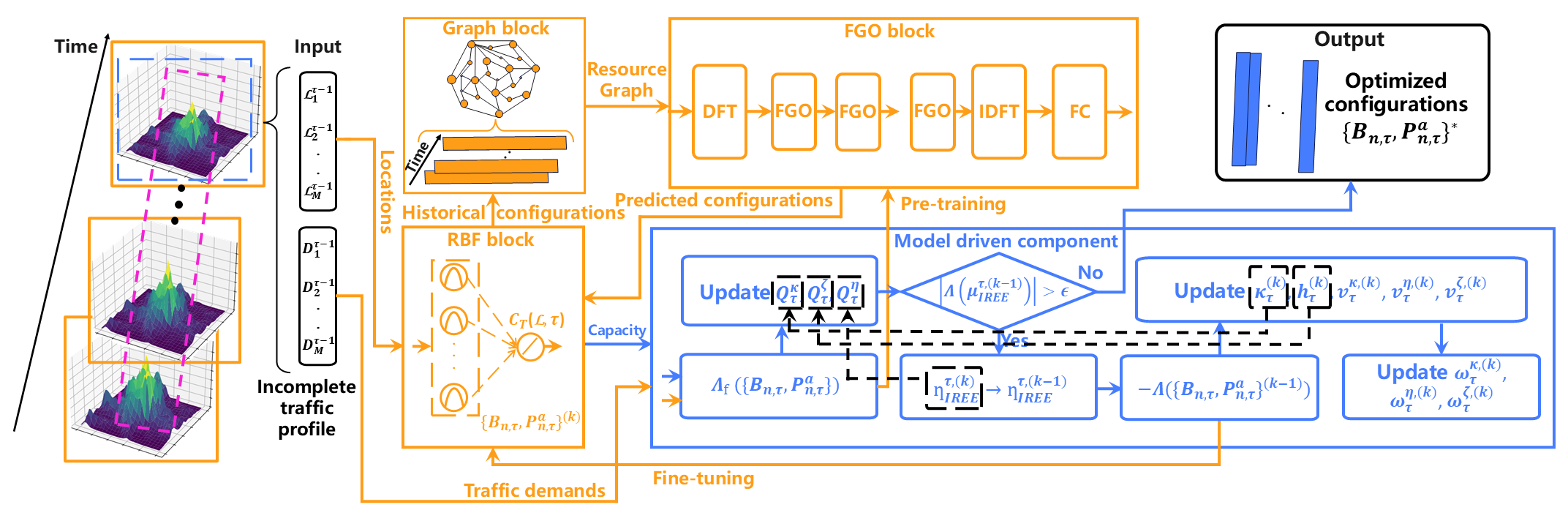}
\caption{Overview of the proposed MDDRA scheme. The orange boxes represent the data-driven component, i.e. the architecture of the proposed GRAF network. The blue box is the model-driven component, which provides long-term corrections and the loss functions for the pre-training \& fine-tuning phases. The black dashed lines indicate the long-term error sources of the system.
}
\label{fig:proposed}
\end{figure*}

\begin{Prob}[Lower Bound Maximization] The lower bound of IREE for the entire wireless network with $N_{BS}$ BSs can be maximized by solving the following optimization problem.
\label{prob:lower bound}
\begin{eqnarray}
    \underset{\{B_{n,\tau},P_{n,\tau}^{a}\}, \kappa}{\textrm{maximize}} && \eta^{\mathcal{T}}_{IREE}(C_{T}, D), \nonumber \\
    \textrm{subject to} && \eqref{eqn:convection-diffusion} - \eqref{eqn:total_pow_def}, \eqref{constraint:max_bandwidth} - \eqref{def:kappa}, \nonumber \\
    && \zeta\left ( \{B_{n,\tau},P_{n,\tau}^{a}\}, \kappa \right ) \geq \zeta_{\min}, \label{constraint:zeta} %\\
    % && \kappa \triangleq \sum^{\mathcal{T}}_{\tau=0} 
    % \xi(\tau) \label{def:kappa}, 
\end{eqnarray}
where $\zeta(\{B_{n,\tau},P_{n,\tau}^{a}\}, \kappa) = \frac{(1 - \kappa) \sum^{\mathcal{T}}_{\tau=0} \min\{C_{Tot}(\tau),D_{Tot}(\tau)\} }{\sum^{\mathcal{T}}_{\tau=0} D_{Tot}(\tau)} $ is the network utility indicator\footnote{Since $\zeta(\{B_{n,\tau},P_{n,\tau}^{a}\}, \kappa)$ is a lower bound of the term on the left side of the inequality \eqref{constraint:qos}, we have replaced the original constraint \eqref{constraint:qos} with a more strict constraint \eqref{constraint:zeta}.}.
\end{Prob}

To address this fractional programming problem, we employ a two-step iterative procedure. In the $k^{th}$ iteration, the first step involves solving a utility maximization problem for a given $\eta_{IREE}^{\mathcal{T}, (k)}$, as formulated in Problem~\ref{prob:utility}. Subsequently, with the obtained value of $C_{Tot}^{(k)} (\tau)$ and $\kappa^{(k)}$, the second step updates the value $\eta_{IREE}^{\mathcal{T}, (k+1)}$, for the next iteration. This iterative process continues until convergence is achieved.

\begin{Prob}[Utility Maximization for Given $\eta^{\mathcal{T}}_{IREE}$] For any $\eta^{\mathcal{T}}_{IREE}$, the utility for the entire wireless network with $N_{BS}$ BSs can be maximized by solving the following optimization problem.
\label{prob:utility}
\begin{eqnarray}
    \underset{\{B_{n,\tau},P_{n,\tau}^{a}\}, \kappa}{\textrm{maximize}} && (1 - \kappa) \sum^{\mathcal{T}}_{\tau=0} \min\{C_{Tot}(\tau),D_{Tot}(\tau)\} \nonumber \\
    && - \eta^{\mathcal{T}}_{IREE} \sum^{\mathcal{T}}_{\tau=0} P_T (\tau) , \nonumber \\
    \textrm{subject to} && \eqref{eqn:convection-diffusion} - \eqref{eqn:total_pow_def}, \eqref{constraint:max_bandwidth} - \eqref{constraint:zeta}. \nonumber
\end{eqnarray}
\end{Prob}

So far, we can achieve the above process by combining the RBF based approach, which is elaborated in detail in our previous work \cite{10605762}, and a traditional distributed ADMM algorithm \cite{yang2022survey}. However, this requires the complete temporal-spatial traffic profile, i.e., $\{D(\mathcal{L}, \tau)\}$, which is in general difficult to obtain in practical systems.

\section{Proposed Model-Data Dual-Driven Resource Allocation Scheme}
\label{sect:proposed_scheme}

In this section, we propose the MDDRA algorithm to deal with the spatial-temporal mismatch and optimize the IREE under incomplete traffic profile. As the data-driven component of the MDDRA, GRAF network are also proposed, along with its architecture and training methodology.

\subsection{Framework Overview of MDDRA}

As shown in Fig.~\ref{fig:proposed}, we propose a model-data dual-driven framework, which consists of two components, including model-driven long-term IREE correction and data-driven short-term traffic prediction \& IREE-focused fine-tuning.

\begin{itemize}
\item{\em{Model-Driven Long-term IREE Correction.}}
To address Problem~\ref{prob:utility} under incomplete temporal traffic, a natural idea is to maximize the transient IREE given by,
\begin{eqnarray}
\label{eqn:transient_iree}
\eta_{IREE}^{\tau} = \frac{(1 - \kappa_{\tau} ) \min \big\{ C_{Tot}(\tau), D_{Tot}(\tau) \big\} }{P_T (\tau)},
\end{eqnarray} 
where $\kappa_{\tau}$ is the transient copy of $\kappa$. The idea is to find a long-term corrected transient solution and simultaneously ensure the convergence. To be more specific, we apply a model-driven Lyapunov framework to impose queue-based long-term constraints on short-term IREE optimization, mitigating the cumulative errors induced by short-term optimization through the rate stability conditions \cite{neely2022stochastic} as given in Fig.~\ref{fig:proposed}.

\item{\em{Data-Driven Short-term Traffic Prediction \& IREE-Focused Fine-tuning.
}}To deal with the incomplete spatial traffic profiles and generate accurate short-term traffic predictions, we design a data-driven GRAF network, along with a corresponding pre-training \& fine-tuning training methodology as given in Fig.~\ref{fig:proposed}. In the pre-training stage, GRAF network can generate accurate short-term traffic predictions under incomplete spatial traffic profiles through its RBF-based interpolation and Fourier Graph Operator (FGO) layers \cite{yi2024fouriergnn}. In the fine-tuning stage, the objective of GRAF network shifts to the long-term corrected IREE maximization.
\end{itemize}

As given in the remainder of this section, since data-driven online temporal traffic prediction helps to accurately forecast traffic patterns and model-driven Lyapunov framework preserves theoretical convergence guarantee of the proposed algorithm, the data-driven component and the model-driven component can work together effectively\footnote{In practical deployments, the number of BSs involved in the resource scheduling is in general limited and the assoicated communication overhead can be controllable as well\cite{andrews2014will}.}.

% \textcolor{blue}{The hybrid method is necessary, since the model-driven Lyapunov queue based long-term correction provides rate stability guarantees by accumulating historical mismatch information and ensuring convergence under ergodic traffic conditions, while the GRAF network acts as a data-driven short-term predictor, adapting to incomplete spatial-temporal traffic profiles through its RBF-based interpolation and FGO layers. Through this framework, the Lyapunov correction guides the GRAF network’s fine-tuning process, enabling it to focus on minimizing transient JS divergence and maintaining alignment with the long-term IREE objective at the same time. }

% model-driven Lyapunov framework repurposes the GRAF network for IREE optimization to obtain the optimal resource configurations $\{B_{n,\tau}, P_{n,\tau}^{a}\} ^{*}$ while its global stability condition corrects long-term errors through queue-based constraint.\footnote{\textcolor{blue}{In practical deployments, the number of BSs involved in the resource scheduling is in general limited and the assoicated communication overhead can be controllable as well\cite{andrews2014will}.}} The above interplay ensures robustness against incomplete data while avoiding dependency on offline training datasets and the need for precise traffic models or statistical priors.

% \textcolor{blue}{Above interplay ensures robustness against incomplete data while avoiding dependency on offline training datasets and the need for precise traffic models or statistical priors. The design details are provided as below.}

\subsection{Model-Driven Long-term IREE Correction}

According to the rate stability conditions of Lyapunov theory\cite{neely2022stochastic}, we propose some queuing-based constraints and reformulate Problem~\ref{prob:utility} as following.

\begin{Prob}[Recursive Utility Maximization for Given $\tau$]
\label{prob:final}
For any given time slot $\tau$, the transient utility for the entire wireless network with $N_{BS}$ BSs can be maximized via the following recursive version.
\begin{eqnarray}
\underset{ \scriptsize \makecell{ \{B_{n,\tau},P_{n,\tau}^{a} \} , \kappa_{\tau}, \\ 
h_{\tau}, \nu^{\zeta}_{\tau}, \nu^{\kappa}_{\tau}, \nu^{\eta}_{\tau} } }{\textrm{maximize}} && \makecell[l] { (1 - \kappa_{\tau})\min\{C_{Tot}(\tau),  D_{Tot}(\tau)\} \\ - \eta_{IREE}^{\tau} P_T (\tau), } \nonumber \\
\textrm{subject to} && \eqref{eqn:convection-diffusion} - \eqref{eqn:total_pow_def}, \eqref{constraint:max_bandwidth} - \eqref{def:kappa}, \nonumber \\
&& (1 - \kappa_{\tau} - h_{\tau}) \min\{C_{Tot}(\tau),D_{Tot}(\tau)\} \nonumber \\
&& + \nu^{\zeta}_{\tau} Q^{\zeta}_{\tau}  = \zeta_{\min} D_{Tot}(\tau), \label{constraint:h_linear} \\
&& \xi(\tau) + \nu^{\kappa}_{\tau} Q^\kappa_{\tau} = \kappa_{\tau} / \mathcal{T}, \label{constraint:js_linear} \\
&& \frac{\eta_{IREE}^{\tau}}{1 - \kappa_{\tau}} 
+ \frac{\nu^{\eta}_{\tau} Q^{\eta}_{\tau}}{P_T(\tau)} = \frac{\eta^{\mathcal{T}}_{IREE}}{1 - \kappa}, \label{constraint:eta_linear} \\
&& \nu^{\zeta}_{\tau}, \nu^{\kappa}_{\tau}, \nu^{\eta}_{\tau} \in [0,2], h_{\tau} \in [0,1], \label{constraint:correction_factor} 
\end{eqnarray}
where $\nu^{\zeta}_{\tau}$, $\nu^{\kappa}_{\tau}$ and $\nu^{\eta}_{\tau}$ are the correction factors. $h_{\tau}$ is the traffic auxiliary variable which reflects how well the traffic requirement \eqref{constraint:zeta} can be met. $Q^{\zeta}_{\tau}$, $Q^\kappa_{\tau}$ and $Q^{\eta}_{\tau}$ are the virtual queues for the recursive error of $\zeta$, $\kappa_{\tau}$ and $\eta_{IREE}^{\tau}$ given by,
\begin{small}
\begin{eqnarray}
\label{eqn:queues}
\left \{
\begin{aligned}
Q^{\zeta}_{\tau} =&  
\sum_{t=0}^{\tau-1} [  (1 - \kappa_{t} - h_{t}) \min\{C_{Tot}(t),D_{Tot}(t)\}  \\
& - \zeta_{\min} D_{Tot}(t) ], \\
Q^\kappa_{\tau} =& \sum_{t=0}^{\tau-1} ( \xi(t)- \kappa_{t}/ \mathcal{T} ), \\
Q^{\eta}_{\tau} =& \sum_{t=0}^{\tau-1} P_T(t) \Big ( \frac{\eta^{t}_{IREE}}{1 - \kappa_{t}} 
- \frac{\eta^{\mathcal{T}}_{IREE}}{1 - \kappa} \Big).
\end{aligned}
\right.
\end{eqnarray}
\end{small}
\end{Prob}

\begin{Lem}
\label{lem:prob_equivalent}
The optimal solution of Problem~\ref{prob:final} will converge to the optimal solution of Problem~\ref{prob:utility}, if  \eqref{constraint:h_linear}, \eqref{constraint:js_linear}, \eqref{constraint:eta_linear} and \eqref{constraint:correction_factor} holds.
\end{Lem}
\IEEEproof Please refer to Appendix~\ref{appendix:prob_equivalent} for the proof.
\endIEEEproof

The reason that Lemma~\ref{lem:prob_equivalent} holds is that the correction factors are constantly correcting $\{B_{n,\tau}, P_{n,\tau}^{a}\}$ and $\kappa_{\tau}$ at each moment, so that the cumulative error of $\zeta$, $\kappa_{\tau}$ and $\eta_{IREE}^{\tau}$, i.e., $Q^{\zeta}_{\tau}$, $Q^\kappa_{\tau} $ and $Q^\eta_{\tau} $ approaches $0$. The distributed augmented Lagrangian function for Problem~\ref{prob:final} is given by\footnote{For simplicity, the convex constraints \eqref{constraint:max_bandwidth}, \eqref{constraint:max_power} and \eqref{constraint:correction_factor}  are omitted.},
\begin{small}
\begin{eqnarray}
\label{eqn:lagrangian_function}
&&\Lambda \big( \{B_{n,\tau}, P_{n,\tau}^{a}\}, \kappa_{\tau}, h_{\tau}, \nu^{\kappa}_{\tau}, \nu^{\eta}_{\tau}, \nu^{\zeta}_{\tau}, \omega^{\kappa}_{\tau}, \omega^{\eta}_{\tau}, \omega^{\zeta}_{\tau} \big ) \nonumber \\
&=& (1-\kappa_{\tau}) \min\left \{ C_{Tot}(\tau) , D_{Tot}(\tau)\right \} -\eta_{IREE}^{\tau} P_{T}(\tau) \nonumber \\ 
&-& \omega^{\kappa}_{\tau} \left (\xi(\tau) + \nu^{\kappa}_{\tau} Q^\kappa_{\tau} - \frac{\kappa_{\tau}}{\mathcal{T}} \right ) - \frac{\rho^{\kappa}}{2} \left (\xi(\tau) + \nu^{\kappa}_{\tau} Q^\kappa_{\tau} - \frac{\kappa_{\tau}}{\mathcal{T}} \right )^{2}  \nonumber \\
&-& \omega^{\eta}_{\tau} \Big [P_T(\tau) \eta_{IREE}^{\tau} - (1 - \kappa_{\tau} ) \Big( \frac{P_T(\tau) \eta^{\mathcal{T}}_{IREE} }{1 - \kappa} - \nu^{\eta}_{\tau} 
Q^{\eta}_{\tau}  \Big) \Big ]  \nonumber \\
&-& \frac{\rho^{\eta}}{2} \Big [P_T(\tau) \eta_{IREE}^{\tau} - (1 - \kappa_{\tau} ) \Big( \frac{P_T(\tau) \eta^{\mathcal{T}}_{IREE} }{1 - \kappa} - \nu^{\eta}_{\tau} 
Q^{\eta}_{\tau}  \Big) \Big ]^{2} \nonumber \\
&-& \omega^{\zeta}_{\tau} \Big [ (1-\kappa_{\tau }- h_{\tau}   )\min\left \{ C_{Tot}(\tau) , D_{Tot}(\tau)\right \} + \nu_{\tau}^{\zeta }Q_{\tau}^{\zeta} \nonumber \\
&-& \zeta_{\min} D_{Tot}(\tau)\Big ]  - \frac{\rho^{\zeta}}{2} \Big [ (1-\kappa_{\tau }- h_{\tau}   ) \nonumber \\
&\times& \min\left \{ C_{Tot}(\tau) , D_{Tot}(\tau)\right \} + \nu_{\tau}^{\zeta }Q_{\tau}^{\zeta} -\zeta_{\min} D_{Tot}(\tau)\Big ]^2, 
\end{eqnarray}
\end{small}
where $\omega^{\kappa}_{\tau}, \omega^{\eta}_{\tau},\omega^{\zeta}_{\tau}$ are the Lagrange dual variables and $\rho^{\kappa}, \rho^{\eta}, \rho^{\zeta} \in \mathbb{R}^+$ denote positive penalty parameters. Please note that in \eqref{eqn:lagrangian_function} we make some equivalent transformations on \eqref{constraint:eta_linear} to ensure the quadratic nature of $\kappa_{\tau}$.

Inspection of the augmented Lagrangian function reveals that $\Lambda$ is quadratic with respect to $\kappa_{\tau}$, $h_{\tau}$, $\nu^{\kappa}_{\tau}$, $\nu^{\eta}_{\tau}$, and $\nu^{\zeta}_{\tau}$. The dual variables $\omega^{\kappa}_{\tau}$, $\omega^{\eta}_{\tau}$, and $\omega^{\zeta}_{\tau}$ can be iteratively updated using standard ADMM techniques \cite{yang2022survey}. However, the non-convexity of the problem complicates the derivation of the optimal $\{B_{n,\tau}, P_{n,\tau}^{a}\}$, which we will discuss in the next section.

% a detailed solution is deferred to Section~\ref{sect:data_driven}.

% Following the solution of Problem~\ref{prob:final}, $\eta_{IREE}^{\tau}$, $\kappa$, and $\eta_{IREE}^{\mathcal{T}}$ require updating. While $\eta_{IREE}^{\tau}$ is directly updated according to its definition in \eqref{eqn:transient_iree}, updating $\kappa$ and $\eta_{IREE}^{\mathcal{T}}$ necessitates complete spatial-temporal information of $\{B_{n,\tau}, P_{n,\tau}^{a}\}$ and $\{D(\mathcal{L}, \tau)\}$, as specified in \eqref{def:kappa} and \eqref{def:long_iree}. To address this, we leverage the ergodic property of traffic patterns and update $\kappa$ and $\eta_{IREE}^{\mathcal{T}}$ using data from a sufficiently long past time period $[\tau-\Delta \mathcal{T}, \tau]$. 

\begin{table}[t]
\setlength{\tabcolsep}{1.5pt} 
  \renewcommand{\arraystretch}{0.4}
\caption{Detailed configurations of the GRAF network. } 
\label{tab:GRAF}
\centering
\footnotesize
\begin{tabular}{c | c | c}
\toprule
\textbf{Block} &  \textbf{Configurations} & \textbf{Size or Value} \\

% \multicolumn{3}{c}{\textbf{Block}} \\ 
%     \textbf{Configurations} & \textbf{Size or Value} & \\
\midrule
\multirow{2}*{Input}
% & \textbf{User location $\mathcal{L}_{m} $} & $N_{sp} \times 2$  \\
& \textbf{$\{{B}_{n,\tau-1-\Delta\mathcal{T}:\tau-1}\}$}  & $N_{BS} \times \Delta\mathcal{T}$ \\
& \textbf{$\{{P}_{n,\tau-1-\Delta\mathcal{T}:\tau-1}^{a}\}$}  & $N_{BS} \times \Delta\mathcal{T}$ \\
%& \textbf{Output Size} & 56 $\times$ 56 & 14 $\times$ 14 \\
\midrule
\multirow{2}*{Graph block}
% & \textbf{User location $\mathcal{L}_{m} $} & $N_{sp} \times 2$  \\
& Size of $\mathbf{X}_{\tau}$  & $N_{BS}\Delta\mathcal{T} \times 2$ \\
& Size of adj. mat. $\mathbf{A}_{\tau}$ & $N_{BS}\Delta\mathcal{T} \times N_{BS}\Delta\mathcal{T}$\\
%& \textbf{Output Size} & 56 $\times$ 56 & 14 $\times$ 14 \\
\midrule
\multirow{5}*{FGO block}
% \multirow{1}*{\textbf{FGO Operater}} 
% & \textbf{User location $\mathcal{L}_{m} $} & $N_{sp} \times 2$  \\
% & Size of Green's kernel & $ 2 \times 2$
% \\
& Num. of FGO layers, $N_{L}$ & $ 3$
\\
& Num. of FC layers & $1$
\\
& Act. fn. of FGO layer & LeakyReLu
\\
& Act. fn. of FC layer & ReLu
\\
\midrule
\multirow{3}*{RBF block}
% \multirow{1}*{\textbf{FGO Operater}} 
% & \textbf{User location $\mathcal{L}_{m} $} & $N_{sp} \times 2$  \\
& Num. of RBF layers & $1$
\\
& Num. of RBF neurons & $N_{BS}$
\\
& Act. fn. of RBF layer & SE-based RBF\cite{10605762}
\\
\midrule
\multirow{3}*{Loss block}
& Loss fn. for pre-training & $\Lambda_{f} \left( \{B_{n,\tau}\},\{P_{n,\tau}^{a}\} \right )$
\\
& Loss fn. for fine-tuning & $- \Lambda \big( \{B_{n,\tau}, P_{n,\tau}^{a}\}\big )$
\\
& Training epoch within each iteration & $10000$
\\
\midrule
Output
& $\{B_{n,\tau}, P_{n,\tau}^{a}\} ^{*}$ & $N_{BS} \times 2$
\\
%& \textbf{Output Size} & 56 $\times$ 56 & 14 $\times$ 14 \\
% \midrule
% \multicolumn{2}{c}{\textbf{Number of FGO layers, $N_{L}$}} &3 \\ 
% \midrule
% \multicolumn{2}{c}{\textbf{Number of FC layers}} &1 \\ 
\bottomrule
\end{tabular}
\end{table}

\IncMargin{1em}
\begin{algorithm} [t] 
\caption{Proposed MDDRA Algorithm at time stamp $\tau$}
\label{alg:MDDRA}
\SetKwInOut{Input}{input}
\SetKwInOut{Output}{output}
\SetKwRepeat{Do}{do}{while}

\Input{ $N_{BS}$, $\alpha$, $\beta$, $\gamma$, $\lambda$, $P^{c}$, $B_{\max}$, $P_{\max}$, $\Delta \mathcal{T}$, $\mathcal{T}$, $\{ \mathcal{L}_n \}_{n=1}^{N_{BS}}$, $\mathbf{X}_{\tau}$, $\left \{ \left [ \mathcal{L}_m^{\tau-1}, D_m^{\tau-1} \right ] \right \}_{m=1}^{N_U^{\tau-1}} $, $\epsilon > 0$. }

\tcp{Data-Driven Traffic Prediction through Pre-training }

\emph{Pre-training.} Update the parameters of FGO block using loss function
$\Lambda_{f}$ \eqref{eqn:loss_func_f} and last traffic data, $\left \{ \left [ \mathcal{L}_m^{\tau-1}, D_m^{\tau-1} \right ] \right \}_{m=1}^{N_U^{\tau-1}}$.

\emph{Update the graph data.} Update the graph $\mathbf{G}_{\tau}$ using current output of the FGO block, $\{B_{n,\tau}, P_{n,\tau}^{a}\}$.

\emph{Prediction.} Obtain the predicted current traffic demands $D(\mathcal{L}, \tau)$ through the GRAF network.

\While{$| \Lambda(\eta^{\tau, (k-1)}_{IREE}) | > \epsilon$} {

\emph{Update transient IREE.} Obtain $\eta^{\tau, (k)}_{IREE}$ according to \eqref{eqn:transient_iree}.

\tcp{Data-Driven IREE maximized Fine-tuning }

\emph{Fine-tuning.} Fine-tune $ \{B_{n,\tau}, P_{n,\tau}^{a}\} $ using the RBF block, $ \{B_{n,\tau}, P_{n,\tau}^{a}\} ^{(k)} = \arg \underset{\{B_{n,\tau}, P_{n,\tau}^{a}\}} {\max} \Lambda \big( \{B_{n,\tau}, P_{n,\tau}^{a}\}\big )$.

\tcp{Model-Driven Long-term Correction}

\emph{Update transient JS divergence.} Update $\kappa_{\tau}^{(k)}$ by $\arg \underset{\kappa_{\tau}} {\max} \; \Lambda \big ( \kappa_{\tau} \big ) $.

\emph{Update traffic auxiliary variable.} Update $h_{\tau}^{(k)}$ by $ \arg \underset{h_{\tau}} {\max} \; \Lambda \big ( h_{\tau} ) $.

\emph{Update correction factors.} Update $\nu^{\kappa, (k)}_{\tau}, \nu^{\eta, (k)}_{\tau},\nu^{\zeta, (k)}_{\tau}$ by $ \arg \underset{\nu^{\kappa}_{\tau}, \nu^{\eta}_{\tau},\nu^{\zeta}_{\tau}} {\max} \; \Lambda \big ( \nu^{\kappa}_{\tau}, \nu^{\eta}_{\tau},\nu^{\zeta}_{\tau} \big ) $.

\emph{Update dual variables.} Obtain $\omega^{\zeta,(k)}_{\tau}$, $\omega^{\kappa,(k)}_{\tau}$ and $\omega^{\eta,(k)}_{\tau}$ according to standard ADMM \cite{yang2022survey}.

% The update rules are given by $\omega^{\zeta}_{\tau} = \omega^{\zeta}_{\tau-1} + \rho^{\zeta}  \Big [(1-\kappa_{\tau }- h_{\tau})\min\left \{ C_{Tot}(\tau) , D_{Tot}(\tau)\right \} + \nu_{\tau}^{\zeta }Q_{\tau}^{\zeta} -\zeta_{\min} D_{Tot}(\tau) \Big ]$,
% $ \omega^{\kappa}_{\tau} = \omega^{\kappa}_{\tau-1} + \rho^{\kappa}  \left (\xi(\tau) + \nu^{\kappa}_{\tau} Q^\kappa_{\tau} - \frac{\kappa_{\tau}}{\mathcal{T}} \right ) $ and
% $\omega^{\eta}_{\tau} = \omega^{\eta}_{\tau-1} + \rho^{\eta}  \Big [P_T(\tau) \eta_{IREE}^{\tau} - (1 - \kappa_{\tau} ) 
% \times \Big( \frac{P_T(\tau) \eta^{\mathcal{T}}_{IREE} }{1 - \kappa} - \nu^{\eta}_{\tau} 
% Q^{\eta}_{\tau} \Big) \Big ]$.

$k = k+1$;
}

\emph{Update virtual queues.} Obtain $Q^{\zeta}_{\tau+1}$, $Q^\kappa_{\tau+1}$ and $Q^{\eta}_{\tau+1}$ according to \eqref{eqn:queues}.

\emph{Update long-term JS divergence and IREE.} Obtain $\kappa$ and $\eta_{IREE}^{\mathcal{T}}$, $\kappa = \sum^{\tau}_{t=\tau - \Delta \mathcal{T}} \xi(t)$, $\eta_{IREE}^{\mathcal{T}} = \frac{(1 - \kappa ) \sum^{\tau}\limits_{t=\tau - \Delta \mathcal{T}} \min \big\{ C_{Tot}(t), D_{Tot}(t) \big\} }{\sum^{\tau}_{t=\tau - \Delta \mathcal{T}} P_T (t)}$.

$\tau = \tau+1$;

\Output {$\{B_{n,\tau}, P_{n,\tau}^{a}\} ^{*}$.}
\end{algorithm}
\DecMargin{1em}

\subsection{Data-Driven Traffic Prediction \& IREE-Focused Fine-tuning}
\label{sect:data_driven}

The architecture of the proposed GRAF network is illustrated in Fig.~\ref{fig:proposed}.
Since the historical network configuration is complete, we focus on these data instead of the traffic data. Specifically, we construct a graph structure to present the dynamic interaction between BSs. Denote $\mathbf{X}_{\tau} = \{ B_{n,t} ,P_{n,t}^{a}\}_{n = 1, t = \tau-1-\Delta \mathcal{T}}^{N_{BS}, \tau-1} \in \mathbb{R}^{N_{BS} \Delta \mathcal{T} \times 2}$ as the nodes of the graph, we define an adjacency matrix $\mathbf{A}_{\tau} \in \mathbb{R}^{N_{BS}\Delta \mathcal{T} \times N_{BS}\Delta \mathcal{T} }$, where $A_{i,j} = 1$ if and only if the two BSs $i$ and $j$ are adjacent and $A_{i,j} = 0$ otherwise. The constructed graph is then given by, 
\begin{eqnarray}
\label{eqn:graf_G}
\mathbf{G}_{\tau} &=& (\mathbf{X}_{\tau}, \mathbf{A}_{\tau}).
\end{eqnarray}

In order to incorporate the adjacency matrix $\mathbf{A}_{\tau}$, we adopt the FGO layer to extract the features with tailored kernels. The tailored Green’s kernel $\phi: [n] \times [n] \rightarrow \mathbb{R}^{2 \times 2}$ is given by $\phi_{i,j} = A_{i,j} \circ W $, where $W \in \mathbb{R}^{2 \times 2}$ is a weight matrix and $\circ$ represents the Hadamard product. The FGO is given as $S = \mathcal{F}(\phi) \in \mathbb{C}^{ N_{BS} \Delta \mathcal{T} \times N_{BS} \Delta \mathcal{T} 
\times 2 \times 2}$, where $\mathcal{F}$ denotes Discrete Fourier Transform (DFT)\footnote{With some common assumptions on the kernel $\phi$, $S$ can be parameterized into a low dimensional complex-valued matrix $\mathbb{C}^{2 \times 2}$ \cite{yi2024fouriergnn}.}. Therefore, we can have following FGO block,
\begin{eqnarray}
\label{eqn:graf_F}
\{B_{n,\tau}, P_{n,\tau}^{a}\}_{n=1}^{N_{BS}} = \mathcal{F}^{-1} \left \{  \sum_{l = 0}^{N_{L}} \psi(\mathcal{F}(\mathbf{X}_{\tau})S_{0:l}(\mathbf{A}_{\tau})+ b_l) \right \},
\end{eqnarray}
where $\psi$ is the activation function, $b_l$ is the complex-valued biases parameter, $N_{L}$ is the number of layers and $S_{0:l}(\mathbf{A}_{\tau}) = \prod_{i = 0}^{l} S_{i}(\mathbf{A}_{\tau})$ with $S_{i}(\mathbf{A}_{\tau})$ the FGO in the $l^{th}$ layer. $\mathcal{F}^{-1}$ is the Inverse Discrete Fourier Transform (IDFT).

Finally, we can obtain the network capacity by passing the bandwidth and transmit power through the RBF layer as\footnote{Actually, what we get from the output of the SE based RBF layer is a lower bound on the network capacity, and \cite{10605762} has shown that the gap is negligible under some training tricks.},
\begin{eqnarray} 
\label{eqn:graf_R}
C_{T}(\mathcal{L}, \tau) = \sum_{n=1}^{N_{BS}} B_{n,\tau} \log_2 \left (1 + \frac{ P_{n,\tau}^{a}/L(\mathcal{L}, \mathcal{L}_n)}{ \sigma^2 B_{\max}} \right).
\end{eqnarray}

The detailed network configuration is given in Table~\ref{tab:GRAF}. Upon receiving the incomplete traffic data, $\left \{ \left [ \mathcal{L}_m^{\tau-1}, D_m^{\tau-1} \right ]  \right \}_{m=1}^{N_U^{\tau-1}}$, the GRAF network is trained to approximate it using the RBF layer. Since the FGO block focus on the dynamic of the network configurations, a complete traffic profile is not required for back propagation. During forward propagation, the GRAF architecture leverages the RBF layer output to effectively interpolate missing traffic data. This approach is justified by the universal approximation property of RBF layer and the spatial continuity of its output.

\begin{Thm}[Universal Approximation Property of GRAF Network] \label{thm:arbitrarily_approx} 
For any spatially continuous, temporally auto-regressive traffic $D$, and any location $\mathcal{L}_m$ defined on $\mathbb{R}^d$, there exists a GRAF network $C_{T}(\mathcal{L}_m, \tau)$ as given in \eqref{eqn:graf_G}, \eqref{eqn:graf_F} and \eqref{eqn:graf_R}, such at for any $\epsilon_1, \epsilon_2, \epsilon_3 > 0$,
\begin{eqnarray} 
\label{eq:}
&&\Vert C_{T}(\mathcal{L}_m, \tau) - D(\mathcal{L}_m, \tau) \Vert_2\nonumber \\
&\leq& \epsilon_1 +  \frac{ N_U^{\tau} }{ \beta \sigma^2 } \left (\frac{ N_{BS} P_{\max} }{B_{\max} }  \epsilon_2 +  \epsilon_3 \right ).
\end{eqnarray}
\end{Thm}
\IEEEproof Please refer to Appendix~\ref{appendix:arbitrarily_approx} for the proof.
\endIEEEproof

%%上述的抽象可以使能RBF网络的任意逼近性。

Based on the above analysis, the graph-based FGO layer can effectively captures the spatial-temporal dependencies among BSs, which serves as a physical abstraction of the network topology. In order to effectively obtain the IREE maximized $\{B_{n,\tau}, P_{n,\tau}^{a}\}$, we employ a pre-training \& fine-tuning method to first capture traffic dynamics and then fine-tune the network to obtain the desired configurations. 

\begin{itemize}
\item{\em Traffic prediction pre-training.} In this stage we minimize the Mean Square Error (MSE) between the network capacity $C_{T}(\mathcal{L}, \tau)$ and the incomplete traffic profile $\left \{ \left [ \mathcal{L}_m^{\tau-1}, D_m^{\tau-1}    \right ]  \right \}_{m=1}^{N_U^{\tau-1}} $ to obtain the bandwidth $\{B_{n,\tau}\}$ and power $\{P_{n,\tau}^{a}\}$. The loss function is given by,
\begin{eqnarray}
\label{eqn:loss_func_f}
\Lambda_{f} \left( \{B_{n,\tau}\},\{P_{n,\tau}^{a}\} \right ) = \sum_{m=1}^{N_U^{\tau}} \left [ C_{T}(\mathcal{L}_m^{\tau}) - D_m^{\tau} \right ]^2.
\end{eqnarray}
Please note that only the parameters of FGO block are trained since there are no trainable parameters in the RBF block for now. After the pre-training process, we update the graph $\mathbf{G}_{\tau}$ using the output of the FGO block $\{B_{n,\tau}\},\{P_{n,\tau}^{a}\} $ to maintain a complete historical dynamics of the network configurations. In practical deployments, the parameters of the FGO block can be trained offline to reduce computational overhead.

\item{\em IREE maximized fine-tuning.} After the pre-training stage, the network configurations $\{B_{n,\tau}, P_{n,\tau}^{a}\}$ is obtained in a capacity-traffic matched manner, hence we only need to fine-tune them to solve Problem~\ref{prob:final}. The loss function is $- \Lambda \big( \{B_{n,\tau}, P_{n,\tau}^{a}\}\big )$ as given in \eqref{eqn:lagrangian_function}. In this stage, $\{B_{n,\tau}, P_{n,\tau}^{a}\}$ is regarded as the trainable parameter in the RBF block, while the parameters of the FGO block are frozen.
\end{itemize}

In Table~\ref{tab:MSE_models}, we compare the prediction performance of the proposed GRAF network with the existing methods. The GRAF network demonstrates significant averaged normalized root mean squared error (ANRMSE) improvements under incomplete traffic data, ranging from $43.1\%$ ($N_U^{\tau}=\frac{3}{4}M$) to $44.1\%$ ($N_U^{\tau}= \frac{1}{2} M$), compared to the best baseline (FourierGNN). Different from traditional multi-variate time-series forecasting approaches, the GRAF network can abstract to the physical network typology, and the main advantages are summarized as below.

\begin{table}[t]
    \centering
    \caption{ANRMSE comparison under different models.}
    \label{tab:MSE_models}
    
    \small
    \setlength{\tabcolsep}{1.5pt} % 减少列间距至 3pt
    
    \begin{tabular}{lrrrrr}
        \toprule
        \multicolumn{2}{c}{\multirow{2}{*}{\diagbox[width=6cm]{\textbf{Models $\&$ Layers}}{\textbf{ANRMSE ($\times 10^{-3}$)}}}} & 
        % \multicolumn{1}{c}{\textbf{Models}} & 
        % \multicolumn{1}{c}{\textbf{Layers}} & 
        % \multicolumn{4}{c}{\textbf{NRMSE ($\times 10^{-4}$) $N_U^{\tau}$}} \\
        \multicolumn{4}{c}{$N_U^{\tau}$} \\
        \cmidrule(lr){3-6}
        \multicolumn{1}{c}{} & 
        \multicolumn{1}{c}{} & 
        \multicolumn{1}{c}{$M$} &
        \multicolumn{1}{c}{$\frac{3}{4} M$} & 
        \multicolumn{1}{c}{$\frac{1}{2} M$} & 
        \multicolumn{1}{c}{$\frac{1}{4} M$}\\
        \midrule
        \multicolumn{1}{c}{\textbf{GRAF}} & 
        \multicolumn{1}{c}{\textbf{\makecell{3 FGO layers \\ + 1 FC layer \\ + 1 RBF layer}}} & 
        \multicolumn{1}{c}{\textbf{1.04}} & 
        \multicolumn{1}{c}{\textbf{3.11}} & 
        \multicolumn{1}{c}{\textbf{6.39}} & 
        \multicolumn{1}{c}{\textbf{10.12}}\\
        \midrule
        \multicolumn{1}{c}{\textbf{FEDformer\cite{zhou2022fedformer}}} & 
        \multicolumn{1}{c}{\textbf{\makecell{2 Conv layers \\ + 1 Attn layer \\ + 2 FC layers}}} & 
        \multicolumn{1}{c}{1.21} & 
        \multicolumn{1}{c}{5.83} & 
        \multicolumn{1}{c}{11.89} & 
        \multicolumn{1}{c}{17.53}\\
        \midrule
        \multicolumn{1}{c}{\textbf{GCN\cite{kipf2016semi}}} & 
        \multicolumn{1}{c}{\textbf{\makecell{4 GC layers \\ + 1 FC layer}}} & 
        \multicolumn{1}{c}{1.49} & 
        \multicolumn{1}{c}{6.39} & 
        \multicolumn{1}{c}{13.03} & 
        \multicolumn{1}{c}{18.53}\\
        \midrule
        \multicolumn{1}{c}{\textbf{FourierGNN\cite{yi2024fouriergnn}}} & 
        \multicolumn{1}{c}{\textbf{\makecell{3 FGO layers \\ + 2 FC layers}}} & 
        \multicolumn{1}{c}{1.14} &
        \multicolumn{1}{c}{5.46} &
        \multicolumn{1}{c}{11.42} &
        \multicolumn{1}{c}{17.37}\\
        \bottomrule
    \end{tabular}
\end{table}

% \begin{table}[t]
%     \centering
%     \caption{Averaged normalized MSE under different models. The adopted simulation parameters are listed in Table~\ref{tab:simu_para}.}
%     \label{tab:MSE_models}
    
%     \small
%     \setlength{\tabcolsep}{3pt} % 减少列间距至 3pt
    
%     \begin{tabular}{lrrrrr}
%         \toprule
%         \multicolumn{1}{c}{\textbf{Models}} & 
%         \multicolumn{1}{c}{\textbf{Layers}} & 
%         \multicolumn{1}{c}{$M$} &
%         \multicolumn{1}{c}{$\frac{3}{4} M$} & 
%         \multicolumn{1}{c}{$\frac{1}{2} M$} & 
%         \multicolumn{1}{c}{$\frac{1}{4} M$}\\
%         \midrule
        % \multicolumn{1}{c}{\textbf{GRAF}} & 
        % \multicolumn{1}{c}{0.1} & 
        % \multicolumn{1}{c}{0.1} & 
        % \multicolumn{1}{c}{0.1} & 
        % \multicolumn{1}{c}{0.1}\\
        % \midrule
        % \multicolumn{1}{c}{\textbf{FEDformer\cite{zhou2022fedformer}}} & 
        % \multicolumn{1}{c}{0.1} & 
        % \multicolumn{1}{c}{0.1} & 
        % \multicolumn{1}{c}{0.1} & 
        % \multicolumn{1}{c}{0.1}\\
        % \midrule
        % \multicolumn{1}{c}{\textbf{GCN\cite{kipf2016semi}}} & 
        % \multicolumn{1}{c}{0.1} & 
        % \multicolumn{1}{c}{0.1} & 
        % \multicolumn{1}{c}{0.1} & 
        % \multicolumn{1}{c}{0.1}\\
        % \midrule
        % \multicolumn{1}{c}{\textbf{FourierGNN\cite{yi2024fouriergnn}}} & 
        % \multicolumn{1}{c}{0.1} & 
        % \multicolumn{1}{c}{0.1} & 
        % \multicolumn{1}{c}{0.1} & 
        % \multicolumn{1}{c}{0.1}\\
%         \bottomrule
%     \end{tabular}
% \end{table}

\begin{figure}[t] %[h]
\centering  
\includegraphics[height=8cm,width=9cm]{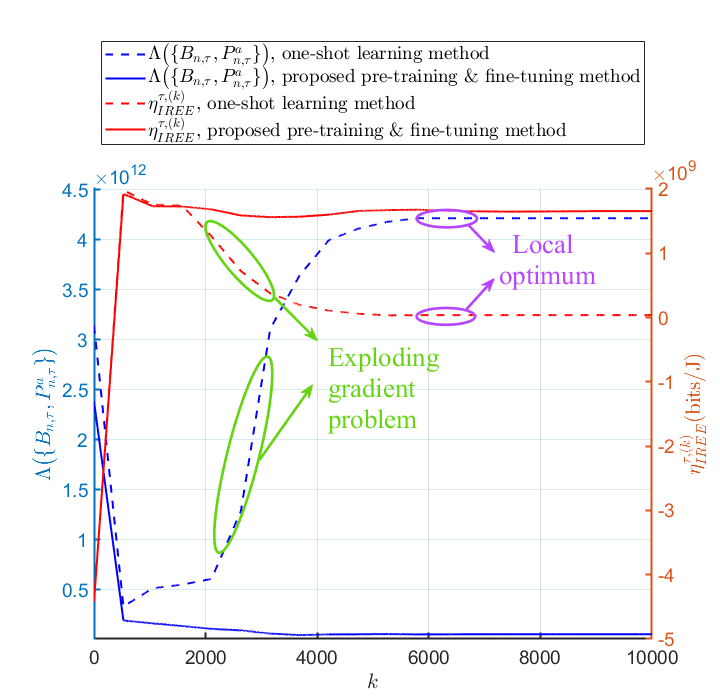}
\caption{Convergence and IREE performance comparison among different training methods. The proposed pre-training \& fine-tuning method deals with mismatch minimization and IREE maximization in different training stages, thus avoiding the exploding gradient problem\cite{rehmer2020vanishing}.}
\label{fig:pre_training_fine_tuning}
\end{figure}

\begin{itemize}
\item{\em Graph Structure versus Non-Graph Structure.}
% \textcolor{blue}{ 
To assess the benefits of incorporating graph structure, we employ FEDformer \cite{zhou2022fedformer}, a frequency domain-based time-series prediction model, as a baseline. While FEDformer effectively captures long- and short-term temporal dynamics through frequency domain analysis, it struggles to capture correlations between variables in multivariate prediction. This limitation hinders its ability to overcome the challenges posed by incomplete spatial traffic data, leading to performance degradation.

% Compare to existing non-graph based time-series prediction models like FEDformer \cite{zhou2022fedformer}

% Traditional time-series prediction models, without graph structures, struggle to effectively capture the spatiotemporal correlations among BSs. By incorporating graph structures, our network accurately models these intricate spatial-temporal relationships, significantly enhancing the extraction of correlations between BSs over time. This results in superior predictive performance, as illustrated in Fig.\ref{fig:4}.}
\item{\em FGO layer versus Graph Convolutional Layer.}
Graph Convolutional Network (GCN) \cite{kipf2016semi} is a classic network architecture that leverage graph structure. While both FGO layers and GCNs update node information by convolving neighbor nodes, the FGO layer assigns varying importance to neighbor nodes at different diffusion steps \cite{yi2024fouriergnn}. This mechanism effectively models scenarios with unbalanced loads among BSs, leading to improved prediction performance.

% A classic network design using graph structure for is Graph Convolutional Network (GCN).
% FGO layer and GCN both update node information via convoluting neighbor nodes. But different from GCN, FGO layer assigns varying importance to neighbor nodes in different diffusion steps as illustrated in \cite{yi2024fouriergnn}, this models the situation of unbalanced load among base stations very well and lead to a better prediction performance.

% . We provide a detailed comparison in Appendix D.

% \textcolor{blue}{ Compared to Graph Convolutional Network (GCN) that operate in the time domain, Frequency-domain Graph Operators (FGO) are better suited for capturing the low-frequency features of dynamic traffic. These features, which are concentrated in the configuration data of BSs, play a critical role in traffic prediction tasks. By focusing on these globally significant patterns, FGOs achieve superior performance in prediction tasks, as demonstrated in Fig. \ref{fig:4}.}
\item{\em RBF Layer versus Fully Connected Layer.}
The conventional neural networks often employ Fully Connected (FC) layers for interpolation due to its universal approximation capabilities  \cite{yi2024fouriergnn}. However, FC layers treat each data point as an independent node, neglecting spatial correlations. Consequently, FC layers struggle to effectively utilize location information when faced with incomplete traffic data.  In contrast, the proposed GRAF network adopts the RBF layer, which not only provides the universal approximation capabilities but also guarantees the spatial continuity of the traffic thanks to the inherent smoothness of radial basis functions.
% While conventional network often employ Fully Connected (FC) layers for interpolation \cite{yi2024fouriergnn}, this approach has limitations. Although FC layers offer universal approximation capabilities, they treat each traffic data point as an independent node, neglecting spatial correlations. Consequently, FC layers struggle to effectively utilize location information when faced with incomplete traffic data.  In contrast, the proposed GRAF network replaces the FC layers with an RBF layer. By leveraging the inherent smoothness of radial basis functions, the RBF layer achieves more accurate interpolation of traffic at adjacent locations, effectively capturing spatial dependencies.
\end{itemize}

Further, the proposed pre-training \& fine-tuning method alleviates the conflict between the objective function and the constraints in Problem~\ref{prob:final}, thereby avoiding the gradient explosion in the one-shot learning method as shown in Fig.~\ref{fig:pre_training_fine_tuning}.

% \subsection{Convergence Properties of MDDRA Scheme} 

\subsection{Convergence Properties and Complexity Analysis}

By combining the proposed pre-training \& fine-tuning method with the long-term iteration, we are able to gradually train the proposed GRAF network for any given IREE value, and iteratively obtain the optimized IREE by tracking both the long-term and the short-term IREE variations. The entire MDDRA scheme has been summarized in Algorithm~\ref{alg:MDDRA}. The convergence properties of the MDDRA scheme can be guaranteed and summarized as below.

\begin{Thm}[Convergence Properties of MDDRA Scheme] \label{thm:conver_analy}
If the loss function $-\Lambda(\eta^{\tau, (k)}_{IREE})$ satisfies the $(L_0,L_1)$ smoothness and $ \lim_{k \rightarrow \infty} \Lambda(\eta^{\tau, (k)}_{IREE}) = 0$, then $\eta^{\mathcal{T}}_{IREE}$ converges to the optimal IREE value\footnote{Although the above convergence analysis relies on idealized traffic assumptions, including the traffic ergodicity and the simplified traffic models, the proposed MDDRA scheme can still converge in many practical scenarios as discussed in Section~\ref{sect:num_res}.}, i.e., $ \lim_{k, \tau \rightarrow \infty} \eta^{\mathcal{T}}_{IREE} = \eta^{\mathcal{T},*}_{IREE}$.
\end{Thm}
\IEEEproof Please refer to Appendix~\ref{appendix:conver_analy} for the proof. 
\endIEEEproof

In the proposed scheme, we require the order of $O(N_{BS} \Delta \mathcal{T} \log ( N_{BS} \Delta \mathcal{T}))$ to pre-train the FGO block at each time stamp \cite{yi2024fouriergnn}, and $O(N_{epoch} N_{ite} ( N_{BS}^2 + M N_{BS}))$ to fine-tuning the RBF layer \cite{10605762}, where $N_{ite}$ is the number of iterations in Dinkelbach's algorithm, and $N_{epoch}$ is the number of epochs. The overall complexity is then given by $O(N_{BS}[ N_{epoch} N_{ite} ( N_{BS} + M ) +  \Delta \mathcal{T}  \log ( N_{BS} \Delta \mathcal{T})  ])$. Note that since we only need to fine-tune the current bandwidth and power to achieve optimized IREE, the number of epochs $N_{epoch}$ can be drastically reduced\footnote{The running time in the fine-tuning stage can be further reduced by using single-instruction multiple-thread GPUs \cite{feng2008multigrid}.} . Meanwhile, for many practical applications, the number of BSs within each control period is in general limited (for example, 10 BSs), and the running time can be further reduced by employing sufficient computing power.

% \begin{table}[t]
% \centering
% \caption{\textcolor{blue}{Running time and achievable IREE performance of the proposed MDDRA Scheme under a single-thread Intel i9-13900HX CPU.}}
% \label{tab:complexity}
% \renewcommand{\arraystretch}{1.5}
% \setlength{\tabcolsep}{14pt} % 调整列间距
% \begin{tabular}{c|c|c}

% \hline
% \textcolor{blue}{$N_{BS}$} & \textcolor{blue}{Running Time (s)} & \textcolor{blue}{\makecell{Achievable \\ IREE ($\times 10^9$ bits/J)} }\\ 
% \hline
% \hline
% \textcolor{blue}{10} & \textcolor{blue}{30.12} & \textcolor{blue}{1.02} \\ 
% \hline
% \textcolor{blue}{20} & \textcolor{blue}{36.04} & \textcolor{blue}{1.15} \\ 
% \hline
% \textcolor{blue}{50} & \textcolor{blue}{52.64} & \textcolor{blue}{1.24} \\ 
% \hline
% \textcolor{blue}{100} & \textcolor{blue}{156.6} & \textcolor{blue}{1.41} \\ 
% \hline
% \textcolor{blue}{200} & \textcolor{blue}{484.6} & \textcolor{blue}{1.45} \\ 
% \hline
% \textcolor{blue}{400} & \textcolor{blue}{1320.3} & \textcolor{blue}{1.47} \\ 
% \hline
% \textcolor{blue}{800} & \textcolor{blue}{3047.9} & \textcolor{blue}{1.48} \\ 
% \hline
% \textcolor{blue}{1200} & \textcolor{blue}{6692.1} & \textcolor{blue}{1.49} \\ 
% \hline
% \end{tabular}
% \end{table}

\section{Numerical Results}
\label{sect:num_res}

\begin{figure*}[t]
\centering\includegraphics[height=5.5cm,width=17.5cm]{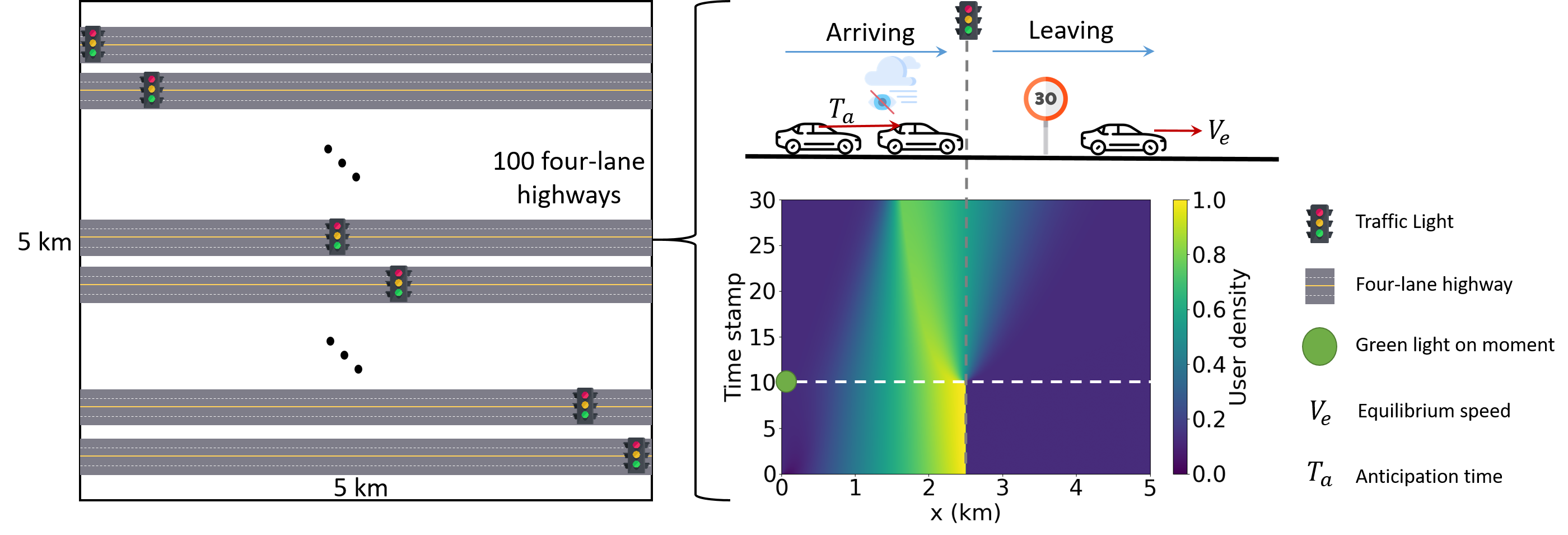}
\caption{ 
An illustration of the simulation scenario, which encompasses an area $\mathcal{A}$ containing $100$ horizontal four-lane roads with traffic signals deployed at equidistant intervals. From time stamp $0$ to $10$, all traffic lights remain red. Subsequently, from time stamp $10$ to $30$, all traffic lights switch to green.
} \label{fig:simu_scenario}
\end{figure*}

In this section, we illustrate the advantages of the proposed MDDRA scheme through some numerical examples. To be more specific, we compare the IREE of the proposed scheme with pure data-driven or model-driven designs to show the effectiveness. By investigating the IREE variation under different traffic dynamics, we conclude with some design principles in time-varying environments.

\begin{table} [htpb] % [h] 
\centering 
\caption{Simulation Settings\cite{10605762,bonzani2000hydrodynamic,kuhne1991macroscopic}}  
\label{tab:simu_para}
\footnotesize
\begin{tabular}{c | c }  

\midrule
Evaluation geographic area, $\mathcal{A}$ & $5 \times 5$ km$^2$ \\

\midrule
Number of BSs, $N_{BS}$ & $400$  \\

\midrule
Maximum Bandwidth, $B_{\max}$ & $36$ GHz \\

\midrule
Maximum Power, $P_{\max}$ & $30$ dBm \\

\midrule  
Circuit power, $P^{c}$ & $5$ W \\

\midrule  
Efficiency of power amplifier, $1/ \lambda $ & $38\%$ \\

\midrule 
Path loss (dB) & $35 +38\log_{10}(d)$ \\

\midrule 
Power spectral density of noise, $\sigma^2$ & $-174$ dBm/Hz \\

\midrule 
Standard deviation of the shadowing, $\chi$ & $10$ dB \\

\midrule 
Minimum traffic requirement, $\zeta_{\min}$ & $0.8$ \\

\midrule 
Empirical exponents, $\mu_1, \mu_2, \vartheta_1, \vartheta_2$ & \makecell{$\mu_1 = 2.05, \mu_2 = 21.11$, \\ $\vartheta_1 = 1, \vartheta_2 = 1$  } \\

\bottomrule
\end{tabular}  
\end{table}  

In the following evaluations, we choose a square area with edge length equal to $5$ kilometers and set the evaluation time $\mathcal{T}$ to $30$ time stamps. In this area, $N_{BS} = 400$ BSs are deployed, whose location are generated according to the typical simulation settings in Table~\ref{tab:simu_para} using the method in \cite{10605762}. The traffic is obtained according to the convection-diffusion model \eqref{def:traffic}, where the initial user density on each lane is established with a linear distribution from the traffic light to its arrival direction. An illustration of the simulation scenario is depicted in Fig.~\ref{fig:simu_scenario}. We formulate the shadowing effects as a zero-mean log-normal distribution with standard deviation $\chi$ \cite{blaszczyszyn2012quality}. The detailed configurations of the proposed GRAF network are listed in Table~\ref{tab:GRAF} and other simulation parameters, unless otherwise specified, are listed in Table~\ref{tab:simu_para}.

\subsection{IREE Comparison with Baselines}

In order to show the effectiveness of the proposed MDDRA scheme, we compare it with the following four baselines. {\em Baseline 1: Distributed ADMM with complete traffic \cite{yang2022survey}}, where we apply distributed ADMM algorithm with all spatial-temporal traffic data available. {\em Baseline 2: Lyapunov based ADMM with current traffic}, where we replace the predicted traffic data with the ground-truth under MDDRA framework. {\em Baseline 3: Greedy scheme with predicted traffic}, where we optimize the transient IREE with predicted traffic obtained by the GRAF network. {\em Baseline 4: Greedy scheme with averaged historical traffic}, where we optimize the transient IREE using averaged historical traffic data. A detailed comparison of the baselines is given in Table~\ref{tab:base_comp}.

\begin{table}[htpb]
\centering 
\caption{Comparison of Baselines}
\label{tab:base_comp}
\begin{tabular}{ccc}
\hline
Schemes & Available traffic data& Model-driven Algorithm \\ 

\hline
Baseline 1 & Complete spatial-temporal data & \begin{tabular}[c]{@{}c@{}}Distributed ADMM \\ + RBF approach\end{tabular}        \\ 

\hline
Baseline 2 & Complete current data  & \begin{tabular}[c]{@{}c@{}}Long-term correction \\ + RBF Network\end{tabular} \\ 

\hline
\begin{tabular}[c]{@{}c@{}}MDDRA \\ Scheme\end{tabular} &\begin{tabular}[c]{@{}c@{}} Predicted current data \\ using GRAF network \end{tabular} & \begin{tabular}[c]{@{}c@{}}Long-term correction \\ + RBF approach\end{tabular} \\ 

\hline
Baseline 3 & \begin{tabular}[c]{@{}c@{}}Predicted current data \\ using GRAF network \end{tabular}  & RBF approach \\ 

\hline
Baseline 4 & Aaveraged  historical data & RBF approach \\ \hline
\end{tabular}
\end{table}

\begin{figure}[t] %[h]
\centering  
\includegraphics[height=7cm,width=9cm]{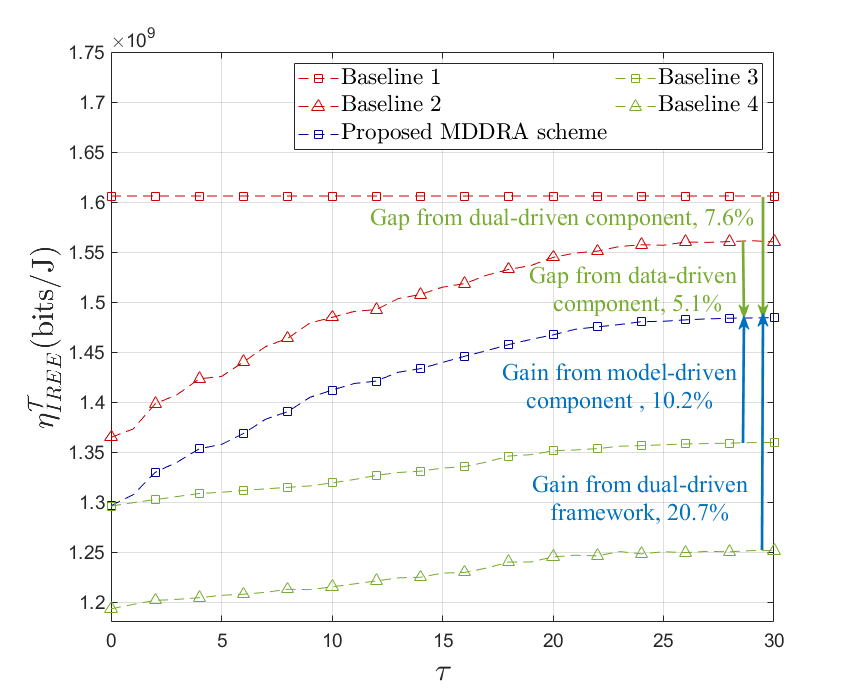}
\caption{ The long-term IREE performance using different schemes. The performance gains and gaps achieved by the proposed dual-driven framework can be observed. } 

\label{fig:effectiveness}
\end{figure}

\begin{figure}[t] %[h]
\centering  
\includegraphics[height=7cm,width=9cm]{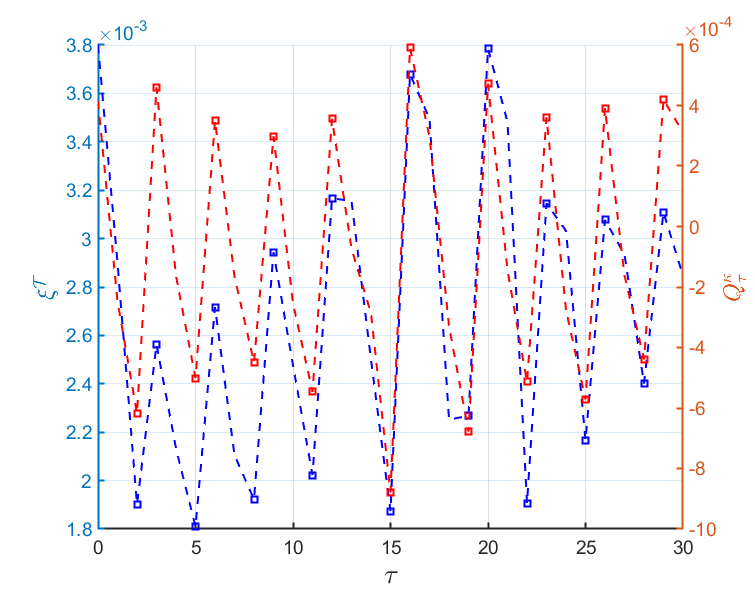}
\caption{ The $\xi^{\tau}$ and $Q_{\tau}^{\kappa}$ dynamics. The virtual queue $Q_{\tau}^{\kappa}$ can effectively capture the changes in $\xi^{\tau}$ caused by the dynamic nature of traffic.}
\label{fig:queue_dynamic}
\end{figure}

As shown in Fig.~\ref{fig:effectiveness}, when compared with Baseline 3, the proposed MDDRA scheme leverages a model-driven Lyapunov framework to compensate for the discrepancy between $\eta_{IREE}^{\tau}$ and $\eta_{IREE}^{\mathcal{T}}$, resulting in a performance improvement of 10.2\%. Furthermore, when benchmarked against Baseline 4, the MDDRA scheme achieves a 20.7\% enhancement in IREE. The additional 10.5\% gain can be attributed to the superior capability of the GRAF network in capturing complex traffic patterns and mitigating the impact of incomplete traffic data. These two comparisons highlight the benefits of integrating model-driven and data-driven methods in the proposed scheme. Given that Baseline 1 and Baseline 2 has access to the complete traffic distributions in time or space, the comparison between proposed MDDRA scheme and them exposes the performance gap. The gap between MDDRA scheme and Baseline 2 is caused by the prediction error of the GRAF network and the gap between MDDRA scheme and Baseline 1 compounds the influence of tracking error caused by the queue dynamics. Notably, even under conditions of incomplete traffic data, these gaps remain within a manageable range, approximately around $5.1\%$ and $7.6\%$. As shown in Fig.~\ref{fig:queue_dynamic}, the virtual queue $Q_{\tau}^{\kappa}$ effectively captures the fluctuations in transient JS divergence $\xi^{\tau}$.

% when comparing Baseline 4 with Baseline 3, we can find that the GRAF network based traffic prediction brings an IREE gain of \textcolor{blue}{$10.5\%$}, which illustrate effectiveness of our data-driven approach in capturing intricate traffic patterns and handling the incompleteness of traffic. While comparing with Baseline 3, proposed MDDRA scheme employs a model-driven Lyapunov framework to correct the error between $\eta_{IREE}^{\tau}$ and $\eta_{IREE}^{\mathcal{T}}$, thereby achieves a improvement of \textcolor{blue}{$10.2\%$}. Above two comparisons highlight the benefits of integrating model-driven and data-driven methods in the proposed MDDRA scheme. Given that Baseline 1 and Baseline 2 has access to the complete traffic distributions in time or space, the comparison between proposed MDDRA scheme and them exposes the performance gap. The gap from MDDRA scheme and Baseline 2 is caused by the prediction error of the GRAF network and the gap between Baseline 2 and Baseline 1 is due to the tracking error caused by the queue dynamics. Notably, even under conditions of incomplete traffic data, these gaps remain within a manageable range, approximately around \textcolor{blue}{$5.1\%$} and \textcolor{blue}{$2.2\%$}. As shown in Fig.~\ref{fig:queue_dynamic}, the virtual queue $Q_{\tau}^{\kappa}$ effectively captures the fluctuations in transient JS divergence $\xi^{\tau}$.

\begin{figure*} [htpb]
\centering
\subfigure[ Network utility and JS divergence versus versus $V_{e}$ under different $\zeta_{\min}$ with $P_{\max}=15$ dbm. ]{
\begin{minipage}[c]{0.45\linewidth}
\centering
\includegraphics[height=7cm,width=8cm]{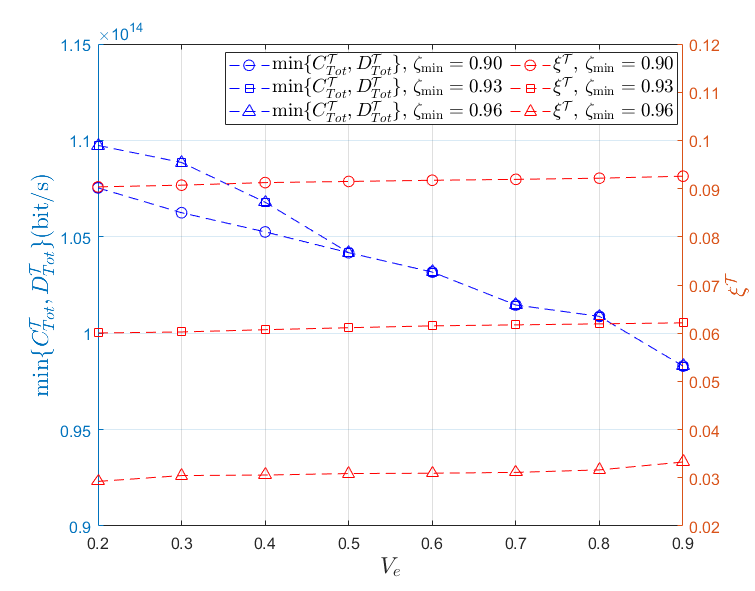}
\label{fig:js_ve}
% \vspace*{3pt}
\end{minipage}}
\subfigure[ Network utility and JS divergence versus $T_{a}$ under different $\zeta_{\min}$ with $P_{\max}=15$ dbm.]{
\begin{minipage}[c]{0.45\linewidth}
\centering
\includegraphics[height=7cm,width=8cm]{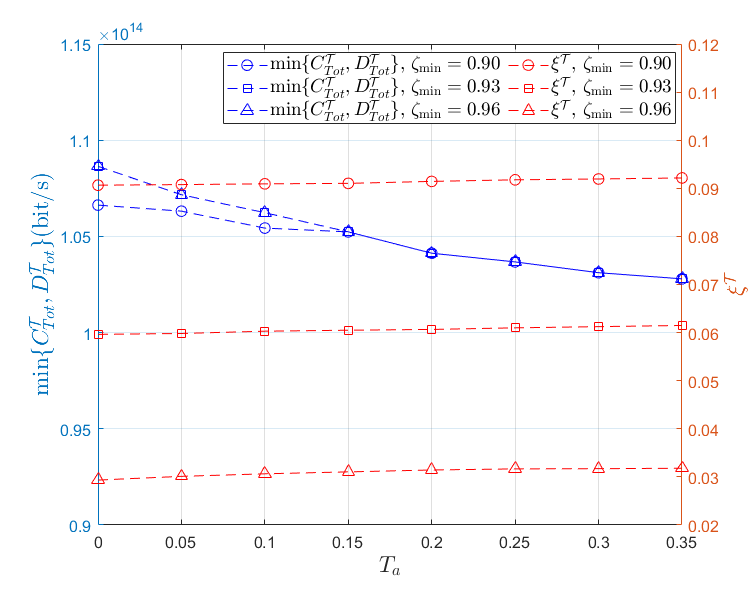}
\label{fig:js_ta}
% \vspace*{3pt}
\end{minipage}}
\subfigure[ IREE/EE versus $V_{e}$ under different $\zeta_{\min}$ with $P_{\max}=15$ dbm. ]{
\begin{minipage}[c]{0.45\linewidth}
\centering
\includegraphics[height=7cm,width=8cm]{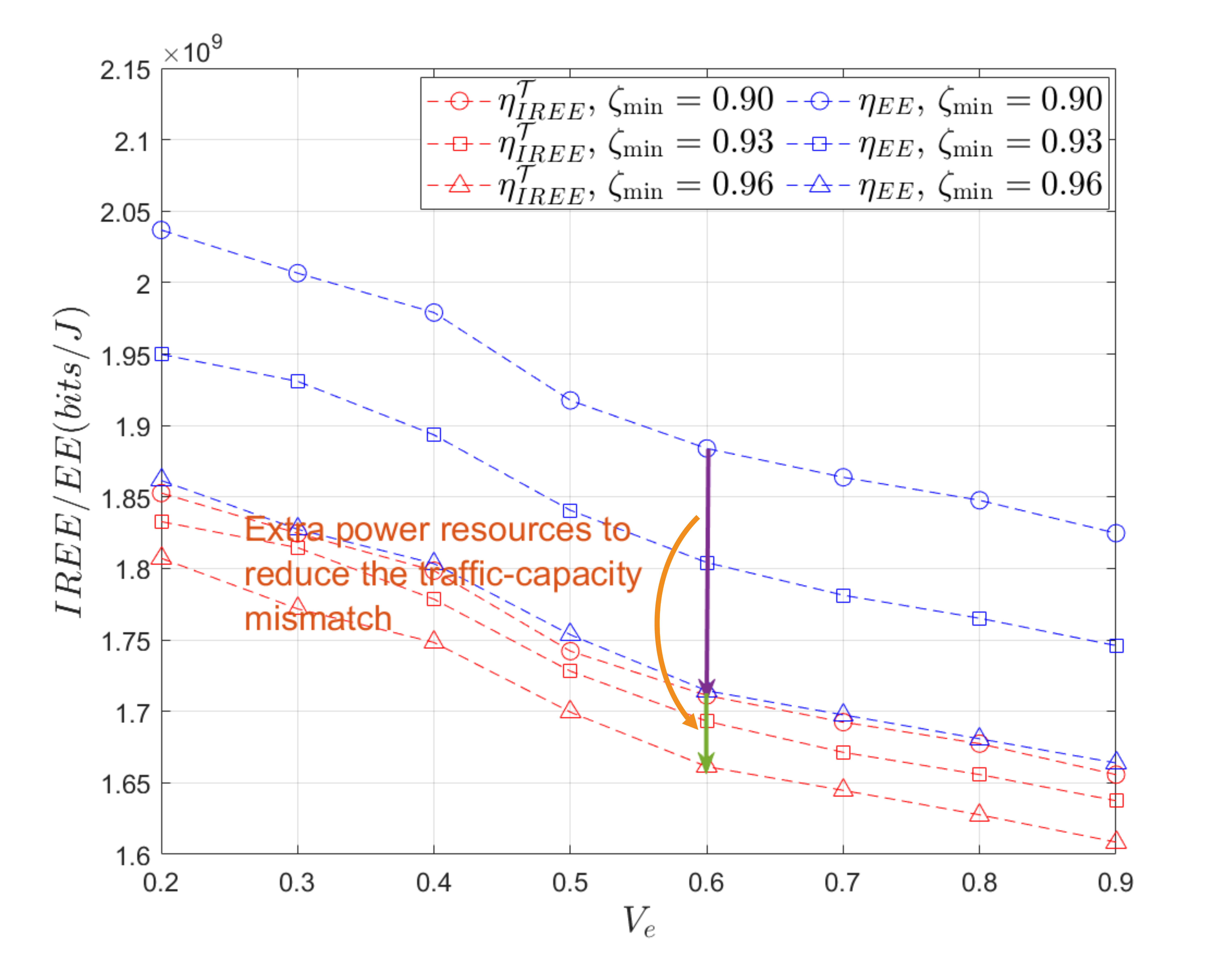}
\label{fig:qos_ve}
% \vspace*{3pt}
\end{minipage}}
\subfigure[ IREE/EE versus $T_{a}$ under different $\zeta_{\min}$ with $P_{\max}=15$ dbm. ]{
\begin{minipage}[c]{0.45\linewidth}
\centering
\includegraphics[height=7cm,width=8cm]{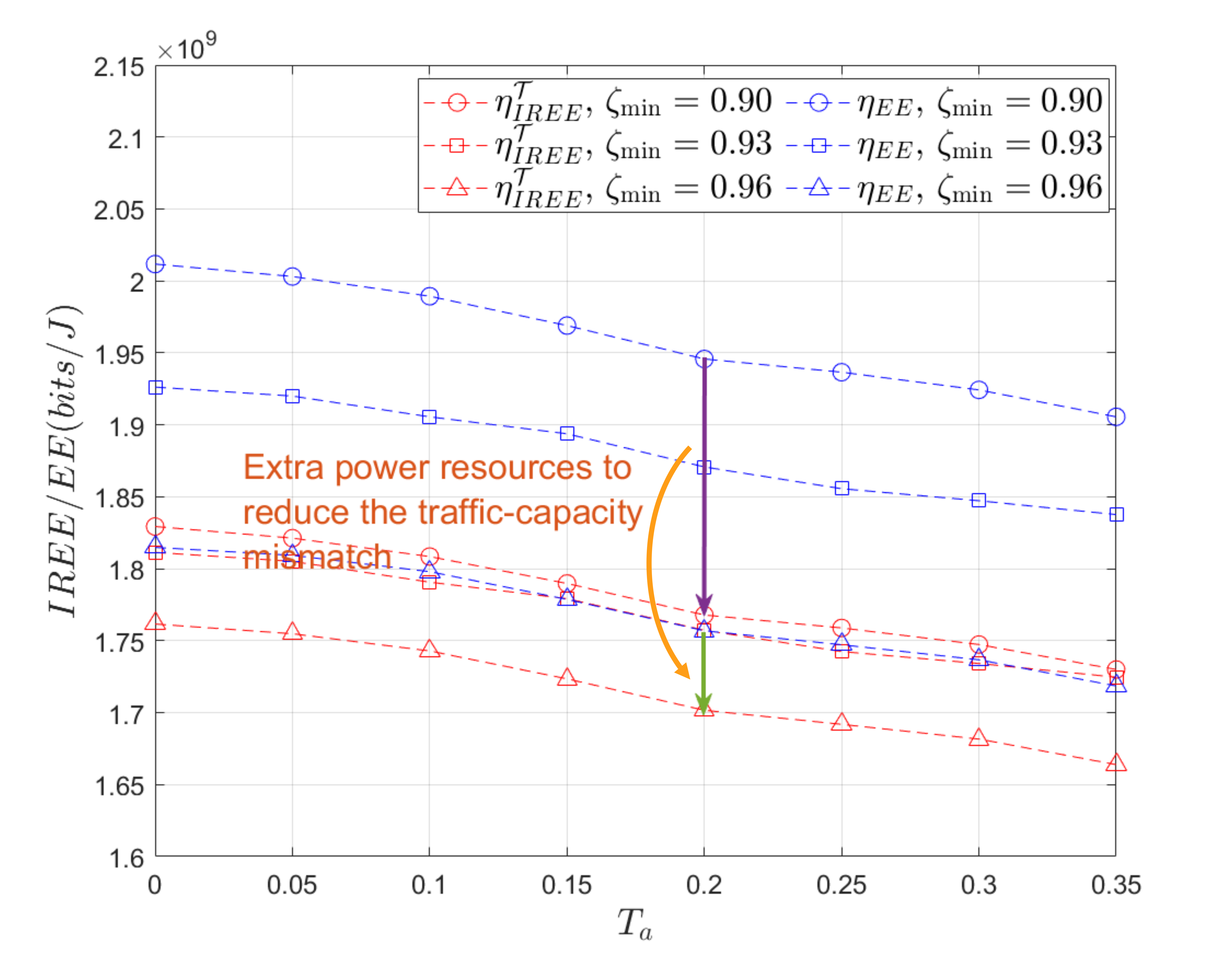}
\label{fig:qos_ta}
% \vspace*{3pt}
\end{minipage}}
\subfigure[IREE/EE versus $V_{e}$ under different $P_{\max}$ with $\zeta_{\min}=0.93$.]{
\begin{minipage}[c]{0.45\linewidth}
\centering
\includegraphics[height=7cm,width=8cm]{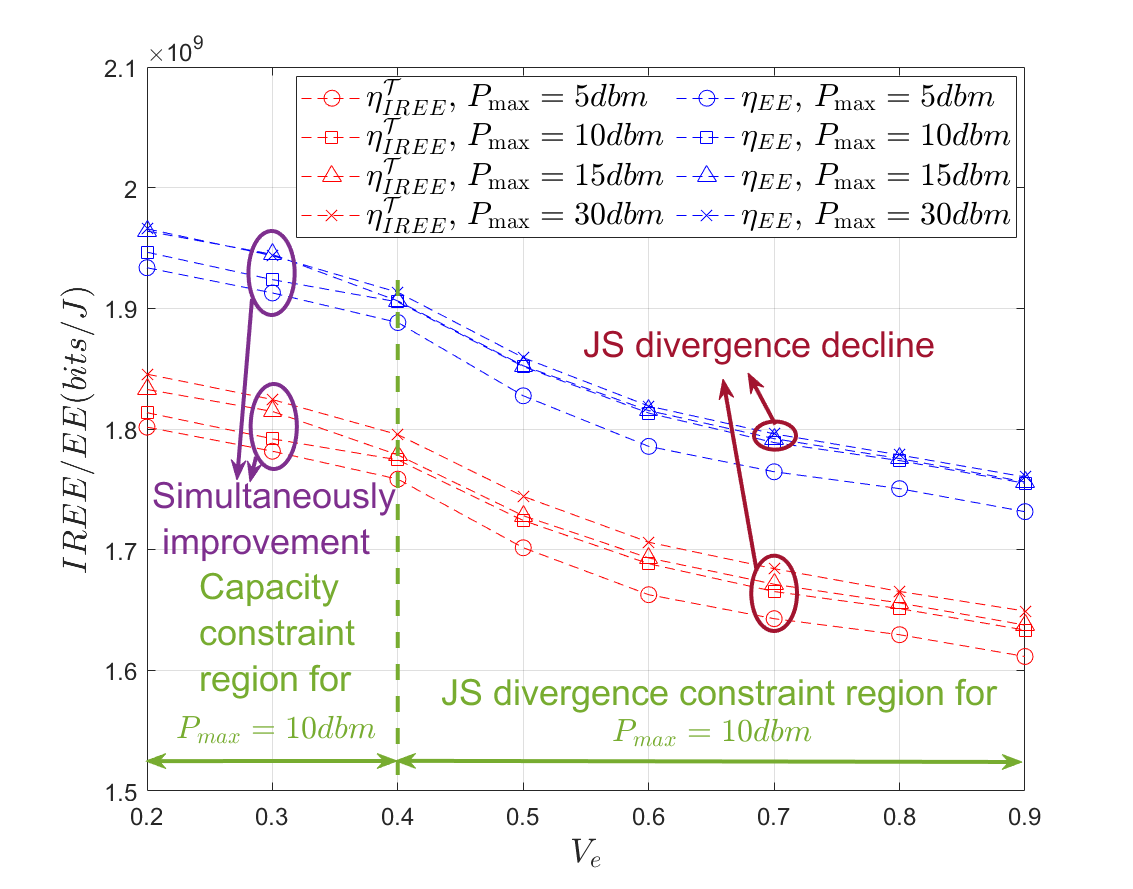}
\label{fig:pmax_ve}
% \vspace*{3pt}
\end{minipage}}
\subfigure[ IREE/EE versus $T_{a}$ under different $P_{\max}$ with $\zeta_{\min}=0.93$.]{
\begin{minipage}[c]{0.45\linewidth}
\centering
\includegraphics[height=7cm,width=8cm]{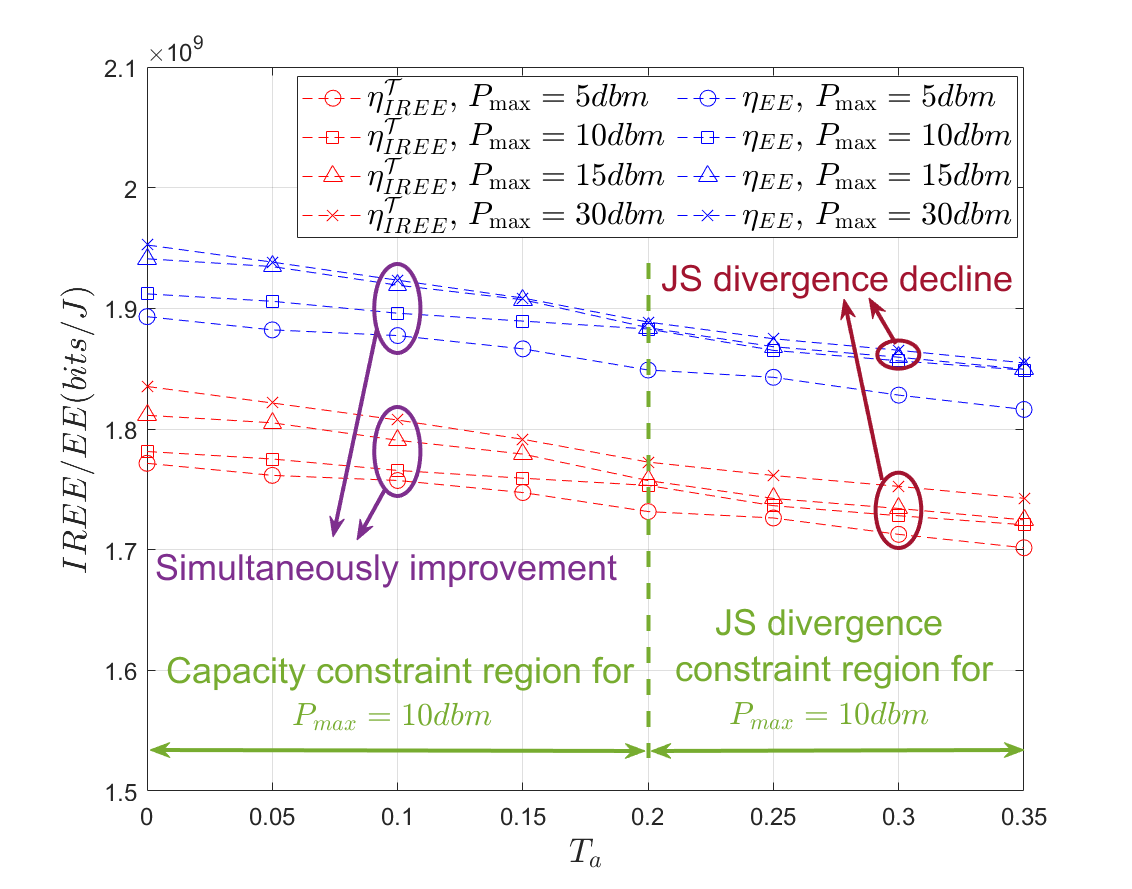}
\label{fig:pmax_ta}
% \vspace*{3pt}
\end{minipage}}
\caption{ Numerical results for the network capacity, JS divergence, EE and IREE versus normalized equilibrium speed and anticipation time. The traffic-capacity mismatch and its impact on IREE under different traffic requirements and power limits can be observed.}
\label{fig:}
\end{figure*}

\subsection{Design Principles}
\label{subsec:design_prin}

Equilibrium speed and anticipation time are fundamental parameters in vehicular flow \cite{newell1961nonlinear}. Equilibrium speed represents the flow speed when the system is in equilibrium, directly influencing traffic density. Anticipation time, conversely, reflects vehicle acceleration and deceleration behavior, as well as inter-vehicle interactions. In real-world scenarios, a road's maximum speed limit directly corresponds to the equilibrium speed \cite{yang2012impact}. Meanwhile, the anticipation time is often associated with driver reaction time and the time required for vehicles to adjust their velocity to a desired speed \cite{kesting2008reaction}, and often exhibits a positive correlation with driving visibility \cite{broughton2007car}. Therefore, we investigate the impact of normalized equilibrium speed $V_{e}$ and normalized anticipation time $T_{a}$ on the traffic-capacity mismatch through numerical simulations in the following.

% Equilibrium speed and anticipation time are among the most important parameters in traffic flow modeling \cite{newell1961nonlinear}. On the one hand, equilibrium speed reflects the number of vehicles passing a point on the road per unit time under a steady state, which has a direct impact on traffic density. On the other hand, expected time reflects the acceleration and deceleration behavior of vehicles and the interaction between vehicles, thereby more realistically simulating the dynamic changes of traffic flow. In real-world systems, the maximum speed limit of a road directly corresponds to the equilibrium speed \cite{yang2012impact}. Furthermore, the anticipation time, another crucial factor, is often related to the reaction time of the drivers and the the velocity adaptation time needed to accelerate to a new desired velocity \cite{kesting2008reaction}. Therefore, in this section, we verify the impact of the normalized equilibrium speed $V_{e}$ and the normalized anticipation time $T_{a}$ on capacity-traffic mismatch, i.e., JS divergence, and IREE performance in a congestion scenario as shown in Fig.~\ref{fig:simu_scenario} through numerical simulation.

\begin{figure}[t] %[h]
\centering  
\includegraphics[height=7cm,width=9cm]{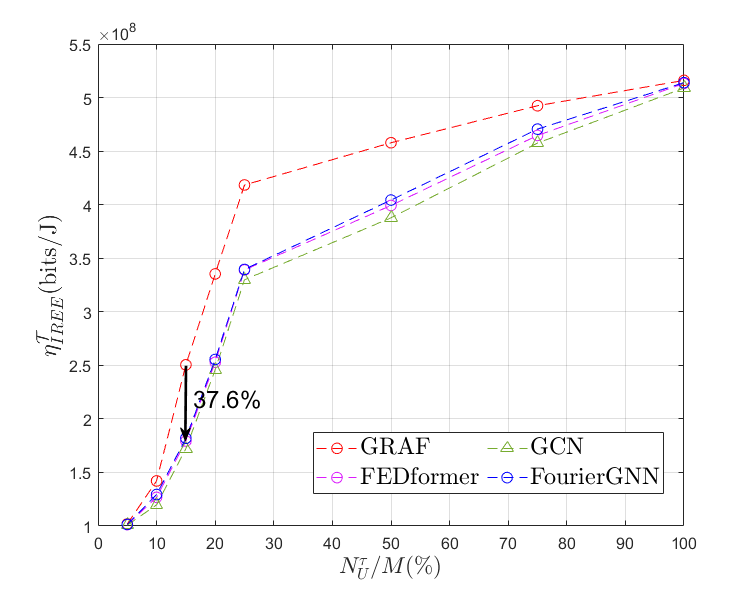}
\caption{ IREE versus $N_{U}^{\tau}/M$ under different data-driven components.
}
\label{fig:robustness}
\end{figure}

\begin{figure}[t] %[h]
\centering  
\includegraphics[height=7cm,width=9cm]{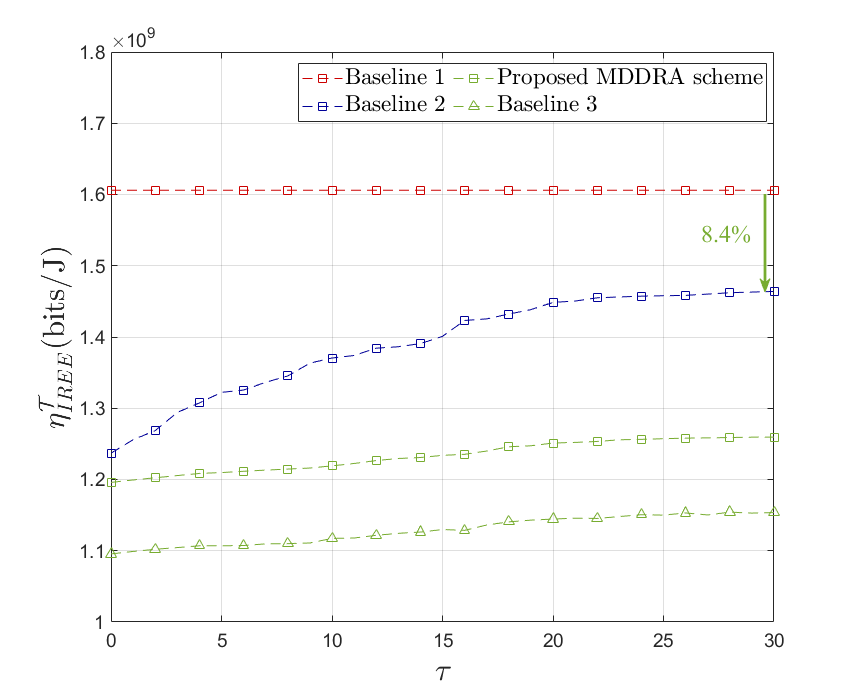}
\caption{The long-term IREE using proposed MDDRA scheme under the real-world map by OpenStreetMap \cite{vargas2020openstreetmap} and measured traffic according to \cite{lopez2018microscopic}.}
\label{fig:sumo_simo}
\end{figure}

With the increase of $V_e$ and $T_a$, the users leave the intersection more quickly. This has two primary consequences as depicted in Fig.~\ref{fig:js_ve} and Fig.~\ref{fig:js_ta}. Firstly, the total traffic demand $D_{Tot}^{\mathcal{T}}$ decreases due to the reduced number of users within the intersection region. Secondly, the JS divergence increases as traffic departs more rapidly from the hotspot area where the majority of base stations are deployed. These two effects result in a decline in IREE and EE performance as shown in Fig.~\ref{fig:qos_ve} and Fig.~\ref{fig:qos_ta}. In addition, as traffic requirement $\zeta_{\min}$ increases and $V_e$ or $T_a$ increases, $\min\{C_{Tot}^{\mathcal{T}},D_{Tot}^{\mathcal{T}}\}$ has to equal to $D_{Tot}^{\mathcal{T}}$ to meet the requirements, which results in the overlap of the $\min\{C_{Tot}^{\mathcal{T}},D_{Tot}^{\mathcal{T}}\}$ under $\zeta_{\min}=0.90$ and $\zeta_{\min}=0.93$ when $V_e \geq 0.7$ or $T_a \geq 0.25$.

We can also observe in Fig.~\ref{fig:qos_ve} and Fig.~\ref{fig:qos_ta} that as the traffic requirement $\zeta_{\min}$ increases, the network allocates more power resources to deal with the traffic-capacity mismatch, hence leading to a reduction in IREE and EE performance. Although the EE performances are reduced due to the extra power consumption, the JS divergence does not show a significant decrease, therefore ensuring the traffic requirement.

Fig.~\ref{fig:pmax_ve} and Fig.~\ref{fig:pmax_ta} illustrate the impact of maximum power constraint $P_{\max}$ on IREE performance. When $P_{\max} = 5$ dbm, the network is primarily constrained by the total system capacity $C_{tot}$. In this region, increasing $P_{\max}$ enhances both EE and IREE. However, when $P_{\max} = 10$ dbm, the limiting factor gradually transitions from $C_{tot}$ to JS divergence as $V_e$ and $T_a$ increase. Consequently, in the capacity constraint region, raising the $P_{\max}$ to $15$ dbm improves both EE and IREE simultaneously.  Conversely, in the JS divergence constraint region, increasing $P_{\max}$ only benefits IREE while EE remains the same, as $C_{tot}$ is already saturated. Similarly, when $P_{\max}$ increases from $15$ dbm to $30$ dbm, EE remains unchanged while IREE continues to improve due to the decline in JS divergence. Therefore, with large speed limits or higher driving visibility (corresponding to larger $V_e$ or $T_a$), extra power budget is required to reduce the traffic-capacity mismatch and ensure the network services even if the total capacity is enough.

\subsection{Extensions}
\label{subsec:ext}
To further illustrate the advantages of the proposed MDDRA scheme, we extend the above simulations in two aspects, robustness and the application to practical scenarios.

In the first experiment, we study the robustness of the proposed MDDRA scheme by varying the sampling rate, e.g., $N^{\tau}_{U}/M$, which aims to measure the IREE performance under different levels of traffic incompleteness. As shown in Fig.~\ref{fig:robustness}, the proposed GRAF network can achieve a maximum IREE gain of $37.6\%$ if compared with GCN \cite{kipf2016semi} or FourierGNN \cite{yi2024fouriergnn} based schemes, since the GRAF structure is capable of providing more reliable traffic predictions.

In the second experiment, we have imported the real-world map using OpenStreetMap \cite{vargas2020openstreetmap} and measured traffic according to \cite{lopez2018microscopic}. Based on the above setting, we re-simulated the above results as summarized in Fig.~\ref{fig:sumo_simo}, where the proposed MDDRA scheme can still converge and outperforming Baseline 3 and Baseline 4 with an optimality gap less than 10\%.

\section{Conclusion} \label{sect:conc}

In this paper, we present an MDDRA framework to deal with the traffic-capacity mismatch and maximize the IREE. By applying a model-driven Lyapunov queue based long-term correction to assemble the historical information and a data-driven traffic prediction GRAF network to characterize the short-time traffic variations under incomplete traffic knowledge, the proposed scheme is able to adapt to the spatial-temporal traffic variations and recursively optimize the wireless resources for performance maximization. The training strategy, the universal approximation property and the convergence properties of the proposed framework are then analyzed. Through some numerical experiments, we validate the performance gains achieved through the data-driven and model-driven components. By analyzing IREE and EE curves under diverse traffic conditions, we recommend that network operators shall pay more attention to balance the traffic demand and the network capacity distribution to ensure the network performance, particularly in scenarios with large speed limits and higher driving visibility.

\begin{appendices}

\section{Derivation of IREE Lower Bound} 
\label{appendix:iree_lower_bound}

Note a function $\xi(\cdot)$ as $\xi(Q_1,Q_2) = \frac{1}{2} \iint_{\mathcal{A}} Q_1 \log_2 \left[ \frac{2 Q_1 }{ Q_1 + Q_2 } \right] + Q_2\log_2 \left[ \frac{2 Q_2 }{ Q_2 + Q_1 }\right] \textrm{d}\mathcal{L} $, we have $\min\{C_{Tot}^{\mathcal{T}},D_{Tot}^{\mathcal{T}}\} \left[1 - \xi^{\mathcal{T}} \right]
= \min\{C_{Tot}^{\mathcal{T}},D_{Tot}^{\mathcal{T}}\} \left[1 - \sum_{\tau=0}^{\mathcal{T}} \xi \left( \frac{C_{T}(\mathcal{L}, \tau)}{C_{Tot}^{\mathcal{T}}}, \frac{D(\mathcal{L}, \tau)}{D_{Tot}^{\mathcal{T}}} \right ) \right] 
\overset{(a)}{\geq} \left[1 - \sum_{\tau=0}^{\mathcal{T}} \xi \left( \frac{C_{T}(\mathcal{L}, \tau)}{C_{Tot}(\tau)}, \frac{D(\mathcal{L}, \tau)}{D_{Tot}(\tau)} \right )\right]  
\min\{C_{Tot}^{\mathcal{T}},D_{Tot}^{\mathcal{T}}\}
\overset{(b)}{\geq} (1 - \kappa ) \sum^{\mathcal{T}}_{\tau=0} \min\{C_{Tot}(\tau),D_{Tot}(\tau)\}$

where the step (a) is obtained due to the convexity of the JS divergence and $\sqrt{\frac{C_{Tot}(\tau)}{C_{Tot}^{\mathcal{T}}} + \frac{D_{Tot}(\tau)}{D_{Tot}^{\mathcal{T}}} } \leq 1$. Step (b) is obtained since $\min\{C_{Tot}^{\mathcal{T}},D_{Tot}^{\mathcal{T}}\} \geq \sum^{\mathcal{T}}_{\tau=0} \min\{C_{Tot}(\tau),D_{Tot}(\tau)\}$. $\kappa = \sum^{\mathcal{T}}_{\tau=0} \xi \left( \frac{C_{T}(\mathcal{L}, \tau)}{C_{Tot}(\tau)}, \frac{D(\mathcal{L}, \tau)}{D_{Tot}(\tau)} \right )$ is the cumulative JS divergence.

\section{Proof of Lemma~\ref{lem:prob_equivalent}}
\label{appendix:prob_equivalent}

For the constraint \eqref{def:kappa}, we can define a queue $Q^\kappa_{\tau} = \sum_{t=0}^{\tau-1} ( \xi(t)- \kappa_{t}/ \mathcal{T} )$ and a correction factor $\nu^{\kappa}_{\tau} \in [0,2]$ such that
\begin{eqnarray}
\label{eqn:}
\xi(\tau) - \kappa_{\tau} / \mathcal{T} = -\nu^{\kappa}_{\tau} Q^\kappa_{\tau}.
\end{eqnarray}
Hence, we have the following Lyapunov drift \cite{neely2022stochastic},
\begin{eqnarray}
\label{eqn:}
\Delta Q^\kappa_{\tau} &=& (Q^\kappa_{\tau+1})^2 - (Q^\kappa_{\tau})^2 =  \big [ Q^\kappa_{\tau} - \nu^{\kappa}_{\tau} Q^\kappa_{\tau} \big]^2 - (Q^\kappa_{\tau})^2 \nonumber \\
&=&  [(1-\nu^{\kappa}_{\tau})^2 - 1](Q^\kappa_{\tau})^2 \leq 0.
\end{eqnarray}
Since $(Q^\kappa_{\tau})^2 \geq 0$, we have $\lim_{\tau \rightarrow \infty} Q^\kappa_{\tau} = 0$, which means that \eqref{def:kappa} is satisfied with large enough $\tau$. Similarly, when equations $(1 - \kappa_{\tau} - h_{\tau}) \min\{C_{Tot}(\tau),D_{Tot}(\tau)\} 
- \zeta_{\min}D_{Tot}(\tau)  = - \nu^{\zeta}_{\tau} Q^{\zeta}_{\tau}$ and $\sum^{\mathcal{T}}_{\tau=0} \Bigg[ \Big ( \frac{\eta_{IREE}^{\tau}}{1 - \kappa_{\tau}} 
- \frac{\eta^{\mathcal{T}}_{IREE}}{1 - \kappa} \Big)    P_T(\tau) \Bigg] = 0$ are satisfied, we have $\lim_{\tau \rightarrow \infty} Q^\zeta_{\tau} = 0$ and $\lim_{\tau \rightarrow \infty} Q^\eta_{\tau} = 0$.

\section{Proof of Theorem~\ref{thm:arbitrarily_approx} }
\label{appendix:arbitrarily_approx}

% From \eqref{eqn:traffic_space}, we know that $D(\mathcal{L}, \tau)$ is continuous. Therefore 

According to \cite{10605762}, there exists a RBF network,
\begin{eqnarray} 
\label{eqn:}
C_{S}(\mathcal{L}, \tau) = \sum_{n=1}^{N_{BS}} \Bar{B}_{n,\tau} \log_2 \left (1 + \frac{ \Bar{P}_{n,\tau}^{a}/L(\mathcal{L}, \mathcal{L}_n)}{ \sigma^2 B_{\max}} \right).
\end{eqnarray}
% \begin{eqnarray}
% \Bar{D} (\{\Bar{B}_{n,\tau}\},\{\Bar{P}_{n,\tau}^{a}\}) &=& \sum_{n=1}^{N_{BS}} \Bar{B}_{n,\tau} \log_2 \Bigg (1 + \nonumber \\
% && \frac{ \Bar{P}_{n,\tau}^{a}/B_{\max} }{ \gamma \sigma^2  \Vert \mathcal{L} - \mathcal{L}_n \Vert_2^{\alpha} + \beta \sigma^2 } \Bigg),
% \end{eqnarray}
such that for all $\mathcal{L}_m$ and any positive $\epsilon_1$ the inequality $\Vert C_{S}(\mathcal{L}_m, \tau) - D(\mathcal{L}_m, \tau) \Vert_2 \leq \epsilon_1$ holds.

Assume that at time stamp $\tau$, we have already trained such RBF network from $\tau - \Delta \tau - 1$ to $\tau - 1$ using the past traffic data $\left \{ \left \{ \mathcal{L}_m^{t}, D_m^{t} \right \}_{m=1}^{N_U^{t}}  \right \}_{t = \tau - \Delta \tau - 1}^{\tau-1}$, and obtained the optimized network parameters $\{ \Bar{B}_{n,t}, \Bar{P}_{n,t}^{a} \}^{\tau-1}_{t = \tau - \Delta\tau-1}$. Since $D(\cdot, \tau)$ defined in \eqref{def:traffic} is a auto-regressive system, we can construct a auto-regressive system for $\Bar{B}_{n,t}, \Bar{P}_{n,t}^{a}$ as follows, 
\begin{eqnarray}
\{\Bar{B}_{n,\tau}, \Bar{P}_{n,\tau}^{a}\} &=& \mathcal{F}_T \big( \{ \Bar{B}_{n,t}, \Bar{P}_{n,t}^{a} \}^{\tau-1}_{t = \tau - \Delta\tau-1} \big).
\end{eqnarray}
For this system, since we have the full historical data, we construct a FGO block to learn this process. Note the output as $\{\hat{B}_{n,\tau}\}$ and $\{\hat{P}_{n,\tau}^{a}\}$, based on the universal approximation theorem \cite{schafer2006recurrent}, we have that for all $\{\hat{B}_{n,\tau}\}, \{\hat{P}_{n,\tau}^{a}\}$ and any positive $\epsilon_2, \epsilon_3$, the following inequality holds,
\begin{eqnarray} 
\label{eq:time_error}
\Vert \hat{B}_{n,\tau} - \Bar{B}_{n,\tau} \Vert_2 \leq \epsilon_2, \quad \Vert \hat{P}_{n,\tau}^{a} - \Bar{P}_{n,\tau}^{a} \Vert_2 \leq \epsilon_3.
\end{eqnarray}
Passing $\hat{B}_{n,\tau}$ and $\hat{P}_{n,\tau}^{a}$ forward through the RBF layer, we can obtain the total output $C_{T}(\mathcal{L}_m, \tau)$. Therefore we have,
\begin{eqnarray} 
\label{eq:}
&&\Vert C_{T}(\mathcal{L}_m, \tau) - C_{S}(\mathcal{L}_m, \tau) \Vert_2   \nonumber \\
&&= \Bigg \Vert \sum_{n=1}^{N_{BS}} \hat{B}_{n,\tau} \log_2 \Bigg (1 + \frac{ \hat{P}_{n,\tau}^{a}/B_{\max} }{ \gamma \sigma^2  \Vert \mathcal{L}_m - \mathcal{L}_n \Vert_2^{\alpha} + \beta \sigma^2 } \Bigg) \nonumber \\
&&-\sum_{n=1}^{N_{BS}} \Bar{B}_{n,\tau} \log_2 \Bigg (1 + \frac{ \Bar{P}_{n,\tau}^{a}/B_{\max} }{ \gamma \sigma^2  \Vert \mathcal{L}_m - \mathcal{L}_n \Vert_2^{\alpha} + \beta \sigma^2 } \Bigg) \Bigg \Vert_2 \nonumber \\
&&= \Bigg \Vert \sum_{n=1}^{N_{BS}} \hat{B}_{n,\tau} \Bigg[ \log_2 \Bigg (1 + \frac{ \hat{P}_{n,\tau}^{a}/B_{\max} }{ \gamma \sigma^2  \Vert \mathcal{L}_m - \mathcal{L}_n \Vert_2^{\alpha} + \beta \sigma^2 } \Bigg) \nonumber \\
&&- \log_2 \Bigg (1 + \frac{ \Bar{P}_{n,\tau}^{a}/B_{\max} }{ \gamma \sigma^2  \Vert \mathcal{L}_m - \mathcal{L}_n \Vert_2^{\alpha} + \beta \sigma^2 } \Bigg) \Bigg ] +\sum_{n=1}^{N_{BS}} (\hat{B}_{n,\tau} \nonumber \\
&&- \Bar{B}_{n,\tau} ) \log_2 \Bigg (1 + \frac{ \Bar{P}_{n,\tau}^{a}/B_{\max} }{ \gamma \sigma^2  \Vert \mathcal{L}_m - \mathcal{L}_n \Vert_2^{\alpha} + \beta \sigma^2 } \Bigg)  \Bigg \Vert_2 \nonumber \\
&&\overset{(a)}{\leq} B_{\max} \Bigg \Vert \sum_{n=1}^{N_{BS}} \Bigg [ \frac{ \hat{P}_{n,\tau}^{a}/B_{\max} }{ \gamma \sigma^2  \Vert \mathcal{L}_m - \mathcal{L}_n \Vert_2^{\alpha} + \beta \sigma^2 } \nonumber \\
&&- \frac{ \Bar{P}_{n,\tau}^{a}/B_{\max} }{ \gamma \sigma^2  \Vert \mathcal{L}_m - \mathcal{L}_n \Vert_2^{\alpha} + \beta \sigma^2 } \Bigg ]  \Bigg \Vert_2 +\Bigg \Vert \sum_{n=1}^{N_{BS}} (\hat{B}_{n,\tau} - \Bar{B}_{n,\tau} ) \Bigg \Vert_2  \nonumber \\
&&\times \sum_{n=1}^{N_{BS}} \log_2 \Bigg (1 + \frac{ \Bar{P}_{n,\tau}^{a}/B_{\max} }{ \gamma \sigma^2  \Vert \mathcal{L}_m - \mathcal{L}_n \Vert_2^{\alpha} + \beta \sigma^2 } \Bigg)  \nonumber \\
&&\overset{(b)}{\leq} B_{\max} \Bigg \Vert \sum_{n=1}^{N_{BS}} \Bigg [ \frac{ \hat{P}_{n,\tau}^{a}/B_{\max} }{ \gamma \sigma^2  \Vert \mathcal{L}_m - \mathcal{L}_n \Vert_2^{\alpha} + \beta \sigma^2 } \nonumber \\
&&- \frac{ \Bar{P}_{n,\tau}^{a}/B_{\max} }{ \gamma \sigma^2  \Vert \mathcal{L}_m - \mathcal{L}_n \Vert_2^{\alpha} + \beta \sigma^2 } \Bigg ]  \Bigg \Vert_2 + \frac{ N_U^{\tau} N_{BS} P_{\max} }{ \beta \sigma^2 B_{\max} }  \epsilon_2 \nonumber \\
&&\overset{(c)}{\leq} \frac{ N_U^{\tau} N_{BS} P_{\max} }{ \beta \sigma^2 B_{\max} }  \epsilon_2 + \frac{ N_U^{\tau} }{ \beta \sigma^2 }  \epsilon_3.
\end{eqnarray}
where step (a) is obtained through the triangle inequality and the properties of the log function. Step (b) and step (c) hold since $\Bar{P}_{n,\tau}^{a} 
\leq P_{\max}$, $\gamma \sigma^2  \Vert \mathcal{L}_m - \mathcal{L}_n \Vert_2 \geq 0$ and \eqref{eq:time_error}. Hence, we have,
\begin{eqnarray} 
\label{eq:}
&&\Vert C_{T}(\mathcal{L}_m, \tau) - D(\mathcal{L}_m, \tau) \Vert_2\nonumber \\
&\leq& \Vert C_{T}(\mathcal{L}_m, \tau) - C_{S}(\mathcal{L}_m, \tau) \Vert_2  \nonumber \\
&&+ \Vert C_{S}(\mathcal{L}_m, \tau) - D(\mathcal{L}_m, \tau) \Vert_2 \nonumber \\
&\leq& \epsilon_1 +  \frac{ N_U^{\tau} }{ \beta \sigma^2 } \left (\frac{ N_{BS} P_{\max} }{B_{\max} }  \epsilon_2 +  \epsilon_3 \right ).
\end{eqnarray}

\section{Proof of Theorem~\ref{thm:conver_analy}}
\label{appendix:conver_analy}

In this section, we first prove the convergence for Problem~\ref{prob:final} and then the convergence of the IREE.

\subsection{Convergence for Problem~\ref{prob:final}}

With a given $\eta_{IREE}$, the first order conditions for an approximate stationary point ($\{B_{n,\tau}, P_{n,\tau}^{a}\}^*$, $\kappa_{\tau}^{*}$, $h_{\tau}^{*}$, $\nu^{\kappa,*}_{\tau}$, $\nu^{\eta,*}_{\tau}$ and $\nu^{\zeta,*}_{\tau}$) are given by,
\begin{eqnarray} 
l_1 &=& \nabla \Lambda(\{B_{n,\tau}, P_{n,\tau}^{a}\}^*), \label{stati:bp_gradient} \\
l_2 &=& - \min\left \{ C_{Tot}(\tau) , D_{Tot}(\tau)\right \} + \frac{1}{\mathcal{T}} \Big [ \omega^{\kappa,*}_{\tau} + \rho^{\kappa} \Big ( \xi(\tau) \nonumber \\
&&+ \nu^{\kappa}_{\tau} Q^\kappa_{\tau} - \frac{\kappa_{\tau}^{*} }{\mathcal{T} } \Big ) \Big ] - \Big( \frac{P_T(\tau) \eta^{\mathcal{T}}_{IREE} }{1 - \kappa} -\nu^{\eta}_{\tau} 
Q^{\eta}_{\tau} \Big) \nonumber \\
&&\times \Bigg [ \omega^{\eta,*}_{\tau} + \rho^{\eta} \Big [P_T(\tau) \eta_{IREE}^{\tau} - (1 - \kappa_{\tau} ) \nonumber \\
&&\times \Big( \frac{P_T(\tau) \eta^{\mathcal{T}}_{IREE} }{1 - \kappa} - \nu^{\eta}_{\tau} 
Q^{\eta}_{\tau}  \Big) \Big ] \Bigg ] + \min \{ C_{Tot}(\tau),  \nonumber \\
&&D_{Tot}(\tau) \} \times [ \omega^{\zeta,*}_{\tau} + \rho^{\zeta} \Big [ (1-\kappa_{\tau }- h_{\tau}   )\nonumber \\
&&\times \min\left \{ C_{Tot}(\tau) , D_{Tot}(\tau)\right \} + \nu_{\tau}^{\zeta,*} Q_{\tau}^{\zeta} \nonumber \\
&&-\zeta_{\min} D_{Tot}(\tau)\Big ]^2, \label{stati:kappa_gradient} \\
l_3 &=& \min \{ C_{Tot}(\tau), D_{Tot}(\tau) \} [ \omega^{\zeta,*}_{\tau} + \rho^{\zeta} \Big [ (1-\kappa_{\tau }\nonumber \\
&&- h_{\tau}   ) \times \min\left \{ C_{Tot}(\tau) , D_{Tot}(\tau)\right \} + \nu_{\tau}^{\zeta,*}Q_{\tau}^{\zeta} \nonumber \\
&&-\zeta_{\min} D_{Tot}(\tau)\Big ]^2,  \label{stati:h_gradient} \\
l_4 &=&  Q^\kappa_{\tau} \left [ \omega^{\kappa,*}_{\tau} + \rho^{\kappa} \left (\xi(\tau) + \nu^{\kappa}_{\tau} Q^\kappa_{\tau} - \frac{\kappa_{\tau}}{\mathcal{T}} \right ) \right], \label{stati:vk_gradient}  \\
l_5 &=& (1 - \kappa_{\tau} ) Q^{\eta}_{\tau} \Big [ \omega^{\eta,*}_{\tau} + \rho^{\eta} \Big [P_T(\tau) \eta_{IREE}^{\tau} \nonumber \\
&&- (1 - \kappa_{\tau} ) \Big( \frac{P_T(\tau) \eta^{\mathcal{T}}_{IREE} }{1 - \kappa} + \nu^{\eta}_{\tau} 
Q^{\eta}_{\tau}  \Big) \Big ] \Big ],   \label{stati:veta_gradient} \\
l_6 &=& Q^\zeta_{\tau} \Big [ \omega^{\zeta,*}_{\tau} + \rho^{\zeta} \Big [ (1-\kappa_{\tau }- h_{\tau}   )\min \{ C_{Tot}(\tau), \nonumber \\
&&D_{Tot}(\tau) \} + \nu_{\tau}^{\zeta }Q_{\tau}^{\zeta} -\zeta_{\min} D_{Tot}(\tau)\Big ] \Big], \label{stati:vzeta_gradient}  
\end{eqnarray} 
%条件用引用就行
such that $\Vert l_i \Vert \leq \epsilon_i$ for $i \in \{1,\dots,6\}$. At the same time, we should satisfy the constraints \eqref{constraint:h_linear},
\eqref{constraint:js_linear} and \eqref{constraint:eta_linear}.

According to \cite{wang2022provable}, the Adam optimizer converges to a bounded region when the loss function $\Lambda$ satisfies $(L_0,L_1)$ smoothness. Hence, after enough training epochs $N_{epoch}$, \eqref{stati:bp_gradient} holds and we have $\Lambda \big( \{B_{n,\tau}, P_{n,\tau}^{a}\}, \kappa_{\tau-1}, h_{\tau-1}, \dots, \omega^{\zeta}_{\tau-1} \big ) \geq \Lambda \big( \{B_{n,\tau-1}, P_{n,\tau-1}^{a}\}, \kappa_{\tau-1}, h_{\tau-1}, \dots, \omega^{\zeta}_{\tau-1} \big )$.

Since the augmented Lagrangian function is quadratic with respect to $\kappa_{\tau}$, $h_{\tau}$, $\nu^{\kappa}_{\tau}$, $\nu^{\eta}_{\tau}$ and $\nu^{\zeta}_{\tau}$,  we can guarantee \eqref{stati:kappa_gradient}, \eqref{stati:h_gradient}, \eqref{stati:vk_gradient}, \eqref{stati:veta_gradient}, \eqref{stati:vzeta_gradient} and have $\Lambda \big( \{B_{n,\tau}, P_{n,\tau}^{a}\}, \kappa_{\tau}, h_{\tau}, \nu^{\kappa}_{\tau}, \nu^{\eta}_{\tau}, \nu^{\zeta}_{\tau}, \dots \big ) \geq \Lambda \big( \{B_{n,\tau}, P_{n,\tau}^{a}\}, \kappa_{\tau-1}, h_{\tau-1}, \nu^{\kappa}_{\tau-1}, \nu^{\eta}_{\tau-1}, \nu^{\zeta}_{\tau-1}, \dots \big )$.

Also, according to the update rule of $\omega^{\kappa}_{\tau}, \omega^{\eta}_{\tau}, \omega^{\zeta}_{\tau}$,  we can easily obtain \eqref{constraint:h_linear} according to \eqref{stati:h_gradient}, obtain \eqref{constraint:eta_linear} according to \eqref{stati:veta_gradient} and obtain \eqref{constraint:js_linear} according to \eqref{stati:kappa_gradient}. Further, we have the following inequality $\Lambda \big( \{B_{n,\tau}, P_{n,\tau}^{a}\}, \kappa_{\tau}, h_{\tau}, \nu^{\kappa}_{\tau}, \nu^{\eta}_{\tau}, \nu^{\zeta}_{\tau}, \omega^{\kappa}_{\tau}, \omega^{\eta}_{\tau}, \omega^{\zeta}_{\tau} \big ) \geq \Lambda \big( \{B_{n,\tau}, P_{n,\tau}^{a}\}, \kappa_{\tau}, h_{\tau}, \nu^{\kappa}_{\tau}, \nu^{\eta}_{\tau}, \nu^{\zeta}_{\tau}, \omega^{\kappa}_{\tau-1}, \omega^{\eta}_{\tau-1}, \omega^{\zeta}_{\tau-1} \big ) $.

On the other hand, $\Lambda$ is clearly bounded from above due to the constraints on the transmit power \eqref{constraint:max_power}, the bandwidth \eqref{constraint:max_bandwidth} and auxiliary variables \eqref{constraint:correction_factor}. Therefore, it converges to a stationary point.

\subsection{Convergence of the IREE}

According to Theorem~3 in \cite{10605762}, if the loss function $-\Lambda(\eta^{\tau, (k)}_{IREE})$ satisfies the $(L_0,L_1)$ smoothness and $ \lim_{k \rightarrow \infty} \Lambda(\eta^{\tau, (k)}_{IREE}) = 0$, $\eta_{IREE}^{\tau}$ will converge to a stationary point $\eta_{IREE}^{\tau,*}$ at each time stamp $\tau$, i.e. ,
\begin{eqnarray}
\eta_{IREE}^{\tau,*} \geq \frac{(1 - \kappa_{\tau} ) \min \big\{ C_{Tot}(\tau),D_{Tot}(\tau) \big\} }{P_T (\tau)} \Bigg |_{\{B_{n,\tau}, P_{n,\tau}^{a}\}}.
\end{eqnarray}
Note $\eta_{IREE}^{\mathcal{T}, \tau}$ as the value of $\eta^{\mathcal{T}}_{IREE}$ at time stamp $\tau$, then we have,
\begin{eqnarray}\label{eqn:}
\label{eqn:}
\eta_{IREE}^{\mathcal{T}, \tau - 1} &=& \frac{(1 - \kappa ) \sum^{\mathcal{T}}\limits_{t=0} \min \big\{ C_{Tot}(t), D_{Tot}(t) \big\} }{\sum^{\mathcal{T}}_{t=0} P_T (t)} \nonumber \\
&\leq& \frac{(1 - \kappa ) \sum^{\mathcal{T}}\limits_{t \neq \tau} \min \big\{ C_{Tot}(t), D_{Tot}(t) \big\} }{\sum^{\mathcal{T}}_{t=0} P_T (t)} \nonumber \\
&+& \frac{(1 - \kappa )  P_T (\tau) }{(1 - \kappa_{\tau} ) \sum^{\mathcal{T}}_{t=0} P_T (t)} \eta_{IREE}^{\tau,*} \nonumber \\
&=& \eta_{IREE}^{\mathcal{T}, \tau}
\end{eqnarray}
Therefore, we create an increasing IREE sequence $\big \{ \eta_{IREE}^{\mathcal{T}, \tau} \big \}$. Further, according to Appendix~\ref{appendix:prob_equivalent}, we can guarantee the optimality of the IREE when $\tau \rightarrow \infty$.

\end{appendices}

% \begin{eqnarray}
% S_{m}(\omega) = \frac{1}{2} \left[M(\omega+\omega_c) + M(\omega - \omega_c) \right]  H(\omega) 
% \end{eqnarray}

\bibliographystyle{IEEEtran}
\bibliography{IEEEfull,main}

% \begin{IEEEbiography}[{\includegraphics[width=1in,height=1.25in,clip,keepaspectratio]{taoyu.eps}}]{Tao Yu}
% received the B.E. and M.E. degrees from the School of Communication and Information Engineering, Shanghai University, in 2018 and 2021, respectively, where he is currently pursuing the Ph.D. degree with the School of Communication and Information Engineering. His research fields include energy-efficient communication networks, machine learning, and deep learning in wireless network.
% \end{IEEEbiography}

\end{document}